\documentclass[12pt,preprint]{aastex631}

\usepackage{natbib}
\usepackage{graphicx}
\usepackage{subfigure}
\usepackage{amsmath,amssymb}
\newcommand{\beginsupplement}{%
        \setcounter{figure}{0}
        \renewcommand{\thefigure}{S\arabic{figure}}%
        \renewcommand{\theHfigure}{S\arabic{figure}}%
     }

\shorttitle{Unveiling the Initiation Route of CMEs through their Slow Rise Phase}
\shortauthors{Xing et al.}

\graphicspath{{./}{figures}}

\begin{document}

\title{Unveiling the Initiation Route of Coronal Mass Ejections through their Slow Rise Phase}

\author{Chen Xing}
\affiliation{School of Astronomy and Space Science, Nanjing University, Nanjing, China}
\affiliation{Sorbonne Université, Observatoire de Paris-PSL, École Polytechnique, IP Paris, CNRS, Laboratoire de Physique des Plasmas, Paris, France}

\author{Guillaume Aulanier}
\affiliation{Sorbonne Université, Observatoire de Paris-PSL, École Polytechnique, IP Paris, CNRS, Laboratoire de Physique des Plasmas, Paris, France}
\affiliation{Rosseland Centre for Solar Physics (RoCS), Institute of Theoretical Astrophysics, Universitetet i Olso, Oslo, Norway}

\author{Xin Cheng}
\affiliation{School of Astronomy and Space Science, Nanjing University, Nanjing, China}

\author{Chun Xia}
\affiliation{School of Physics and Astronomy, Yunnan University, Kunming, China}

\author{Mingde Ding}
\affiliation{School of Astronomy and Space Science, Nanjing University, Nanjing, China}

\begin{abstract}
Understanding the early evolution of coronal mass ejections (CMEs), in particular their initiation, is the key to forecasting solar eruptions and induced disastrous space weather. Although many initiation mechanisms have been proposed, a full understanding of CME initiation, which is identified as a slow rise of CME progenitors in kinematics before the impulsive acceleration, remains elusive. Here, with a state-of-the-art thermal-magnetohydrodynamics simulation, we determine a complete CME initiation route in which multiple mainstream mechanisms occur in sequence yet are tightly coupled. The slow rise is first triggered and driven by the developing hyperbolic flux tube (HFT) reconnection. Subsequently, the slow rise continues as driven by the coupling of the HFT reconnection and the early development of torus instability. The end of the slow rise, i.e., the onset of the impulsive acceleration, is induced by the start of the fast magnetic reconnection coupled with the torus instability. These results unveil that the CME initiation is a complicated process involving multiple physical mechanisms, thus being hardly resolved by a single initiation mechanism.
\end{abstract}


\section{Introduction}

The Sun frequently produces violent plasma eruptions such as coronal mass ejections (CMEs), which are released into the interplanetary space and induce geomagnetic storms \citep{Gosling1993} and damage aerospace equipment, satellite communications and power grids \citep{Elovaara2007} when colliding with the Earth's magnetosphere. The forecast of CMEs is thus extremely important for preventing space disasters and exploring habitable exoplanets \citep{Khodachenko2007}.

The evolution characteristics of CMEs and their pre-eruptive structures (e.g. sigmoids and filaments \citep{Cheng2017}; hereafter named as CME progenitors \citep{Chen2011}) in kinematics at various stages lay the foundation for CME predictions. For hours to days before the eruption, CME progenitors are stable and evolve quasi-statically, rising with a small velocity ($< 1\ \textup{km}\ \textup{s}^{-1}$) and tiny acceleration \citep{Xing2018}. In contrast, during the fast eruption, CMEs show an impulsive increase in velocity (up to hundreds to thousands of $\textup{km}\ \textup{s}^{-1}$) and acceleration (up to hundreds of $\textup{m}\ \textup{s}^{-2}$) in a short time (tens to hundreds of minutes) \citep{Zhang2006}. These two stages of kinematics are named as the quasi-static phase and the impulsive acceleration phase, respectively \citep{Zhang2001,Zhang2006,Xing2018}. The understanding of the kinematics during these two phases is relatively thorough, where the former is believed to be a response to the photospheric motion considering their similar characteristic speeds \citep{Yang2004,Kaithakkal2019} and the latter is driven by the coupling of magnetohydrodynamics (MHD) instabilities \citep{Kliem2006,Aulanier2014,Green2018,Aulanier2021} and fast magnetic reconnection \citep{Lin2000,Zhang2006,Jiang2021}.

The initiation of CMEs refers to the transition between the quasi-static state of CME progenitors and the impulsively eruptive state of CMEs. In theoretical works, it usually corresponds to a \textit{turning point} at the start of the impulsive acceleration phase, on the premise that the evolution of CME progenitors and CMEs is composed of the two phases mentioned above. Such a CME initiation is mainly explained by three mainstream mechanisms which are able to break the stable equilibria of CME progenitors (that is, to trigger the CME eruption) \citep{Aulanier2014,Aulanier2021}, i.e., (1) the torus instability \citep{Kliem2006,Zuccarello2015} which refers to a pre-eruptive flux rope losing its equilibrium when the background constraint field decreases rapidly enough with increasing altitude, (2) the breakout reconnection \citep{Antiochos1999,Masson2019} in which the external reconnection at a null point above the pre-eruptive structure removes the constraint of the overlying field, and (3) the tether-cutting reconnection \citep{Moore2001,Jiang2021} in which the CME is caused by the fast reconnection with outflow jets.

However, some observational works demonstrate that CME progenitors actually always leave their quasi-static phase and experience a slow rise phase in a few to tens of minutes prior to the impulsive acceleration phase \citep{Kahler1988,Zhang2001,Cheng2020,Prasad2023}, specifically, manifesting as slowly rising with an increasing velocity (tens of $\textup{km}\ \textup{s}^{-1}$) and a moderate acceleration (tens of $\textup{m}\ \textup{s}^{-2}$) in this period \citep{Cheng2020}. Meanwhile, there is always a weak soft X-ray enhancement as the precursor of the main flare \citep{Simnett1985,Zhang2001} and the CME progenitors often appear as high-temperature structures (known as hot channels) in extreme ultraviolet (EUV) bands \citep{Cheng2012,Zhang2012}. These observations indicate that the CME initiation, as the transition between the quasi-static state of CME progenitors and the impulsively eruptive state of CMEs, is actually a \textit{process} before the impulsive acceleration phase, in which various precursors including the slow rise occur.

Different conjectures have been proposed in observations to understand the CME initiation as a process. On the one hand, it is argued that the eruption-trigger mechanism (e.g., torus instability, breakout reconnection, and tether-cutting reconnection) occurs at the start of the initiation process and drives the slow rise during the initiation \citep{Zhang2001,Zhang2006}, considering the necessary early development of the eruption-trigger mechanism before inducing significant acceleration of CMEs (e.g., \citeauthor{Schrijver2008} \citeyear{Schrijver2008}). On the other hand, it is suggested that the CME initiation is related to the weak magnetic reconnection prior to the onset of eruption-trigger mechanisms, as the magnetic reconnection builds up the CME progenitor and reduces the overlying field restraining it (\citeauthor{Sterling2007} \citeyear{Sterling2007}; \citeauthor{Savcheva2012} \citeyear{Savcheva2012}; \citeauthor{Cheng2020} \citeyear{Cheng2020}; also see the review by \citeauthor{Green2018} \citeyear{Green2018}). Specifically, \cite{Cheng2023} shown that the slow rise and the heating of CME progenitors are caused by a moderate precursor reconnection above the top of pre-flare loops. However, it is obvious that these two types of interpretations are disjointed, indicating that the initiation of a CME event is likely more complex than those described in above-mentioned conjectures and thus any single mechanism may only reveal a part of nature of the CME initiation. To determine a complete understanding of the CME initiation and to reveal how different physics are coupled with each other, we here perform a state-of-the-art three-dimensional (3D) observationally-inspired thermal-MHD simulation. The comprehensive analyses disclose a complete CME initiation that involves multiple physical mechanisms and even a coupling between two mechanisms.

\section{3D Observationally-inspired Thermal-MHD Simulation of CME}
\subsection{Equations}
In this work, we study the CME initiation by performing a 3D observationally-inspired thermal-MHD simulation labeled as ``Simulation Driven-eruption'' (abbreviated as ``Simulation De'') with the code MPI-AMRVAC \citep{Xia2018}, which solves the following equations in Cartesian coordinates:
\begin{equation}\label{eq1}
\frac{\partial\rho}{\partial t}+\nabla\cdot(\rho\boldsymbol{v}) = 0
\end{equation}
\begin{equation}\label{eq2}
\begin{aligned}
\frac{\partial(\rho\boldsymbol{v})}{\partial t}+\nabla\cdot[\rho\boldsymbol{v}\boldsymbol{v}+(p+\frac{\boldsymbol{B}^2}{2\mu_0})\boldsymbol{I}-\frac{\boldsymbol{B}\boldsymbol{B}}{\mu_0}] = \rho\boldsymbol{g}+2\mu\nabla\cdot[\boldsymbol{S}-\frac{1}{3}(\nabla\cdot\boldsymbol{v})\boldsymbol{I}]
\end{aligned}
\end{equation}
\begin{equation}\label{eq3}
\frac{\partial \boldsymbol{B}}{\partial t}+\nabla\cdot(\boldsymbol{v}\boldsymbol{B}-\boldsymbol{B}\boldsymbol{v}+\psi\boldsymbol{I}) = -\nabla\times(\eta\boldsymbol{J})
\end{equation}
\begin{equation}\label{eq4}
\begin{aligned}
\frac{\partial e_{\textup{int}}}{\partial t}+\nabla\cdot(e_{\textup{int}}\boldsymbol{v})=-p\nabla\cdot\boldsymbol{v}+2\mu[\boldsymbol{S}:\boldsymbol{S}-\frac{1}{3}(\nabla\cdot\boldsymbol{v})^2]+\eta J^2+\nabla\cdot[\kappa_{||}(\boldsymbol{b}\cdot\nabla T)\boldsymbol{b}]
\end{aligned}
\end{equation}
\begin{equation}\label{eq5}
\nabla\times\boldsymbol{B} = \mu_0\boldsymbol{J}
\end{equation}
\begin{equation}\label{eq6}
\frac{\partial\psi}{\partial t}+c_h^2\nabla\cdot\boldsymbol{B} = -\frac{c_h^2}{c_p^2}\psi,
\end{equation}
where $\rho$, $p$, $T$, $e_{\textup{int}}$, $\boldsymbol{v}$, $\boldsymbol{B}$, $\boldsymbol{J}$ represent the mass density, thermal pressure, temperature, internal energy, velocity, magnetic field and current density, respectively. $\boldsymbol{g}=-g\boldsymbol{e}_z$ is the gravity acceleration. $\mu$ is the dynamic viscosity coefficient and set to $10^{-4}$ (in dimensionless unit) throughout the simulation, $\boldsymbol{S}$ is the strain tensor where $S_{ij}=(1/2)(\partial_{i}v_j+\partial_{j}v_i)$, and $\boldsymbol{I}$ is the unit tensor. $\eta$ is the resistivity coefficient whose value is set differently in different phases. $\kappa_{||}=8\times 10^{-7}T^{5/2}$ erg cm$^{-1}$ s$^{-1}$ K$^{-1}$ is the parallel conductivity coefficient, and $\boldsymbol{b}=\boldsymbol{B}/B$ is the normalized magnetic field. We introduce a generalized Lagrange multiplier $\psi$ to maintain the $\nabla\cdot\boldsymbol{B}=0$ condition (i.e., generalized Lagrange multiplier (GLM) method; \citeauthor{Dedner2002} \citeyear{Dedner2002}). The evolution of $\psi$ follows Equation \ref{eq6} and $c_h$ and $c_p$ are constants. As a thermal-MHD simulation, we consider compression heating, viscous dissipation, Ohmic dissipation, and thermal conduction in the energy equation. Especially, we solve the internal energy equation instead of total energy equation to avoid the heating from the numerical resistivity and numerical viscosity.

\subsection{Numerical Methods}
The equations are solved with the HLL scheme, the third-step Runge-Kutta time discretization method, and the fifth-order mp5-limited reconstruction \citep{Suresh1997}. We also use the magnetic field splitting method \citep{Tanaka1994,Xia2018} in this simulation: the magnetic field $\boldsymbol{B}$ is split into two parts, the invariable background field $\boldsymbol{B_0}$ and the deviation $\boldsymbol{B_1}$. In this simulation, $\boldsymbol{B_0}$ is set as the initial magnetic field, and thus the initial $\boldsymbol{B_1}$ is 0. This method is conducive to eliminating numerical errors related to the divergence of magnetic field and achieving more precise results, especially in regions where the magnetic field evolves with minor change from the initial value, although it is relatively less effective in regions where the change of magnetic field is considerable (e.g., the centers of polarities during the converging phase).

The equations are solved in the dimensionless form, with the magnetic permeability equal to 1. The atmosphere is set to a fully ionized ideal gas with a hydrogen-helium abundance ratio of 10:1 \citep{Xia2012}. The dimensionless unit of the length, time, mass density, thermal pressure, temperature, velocity and magnetic field strength is 10 Mm, 67.89 s, $2.34\times10^{-15}$ g cm$^{-3}$, 0.51 erg cm$^{-3}$, 1.6 MK, 147.30 km s$^{-1}$ and 2.53 G, respectively.

The physical domain of the simulation is in the range of $-7\le x\le7$, $-7\le y\le7$, and $0\le z\le14$ (in dimensionless unit). We set a stretched grid ($nx\times ny\times nz=144\times144\times96$),  which is symmetric stretched in $x$ and $y$ directions with stretched ratio of 1.029, and unidirectional stretched in $z$ direction with stretched ratio of 1.028. The spatial resolution in three directions is in the range of 0.0297 $\le dx \le$ 0.2262, 0.0297 $\le dy \le$ 0.2262, and 0.0298 $\le dz \le$ 0.4103 (in dimensionless unit). In dimensional unit, the finest resolutions in three directions are about 300 km, comparable to those of \textit{Solar Dynamics Observatory}/Atmospheric Imaging Assembly (AIA; \citeauthor{Lemen2012} \citeyear{Lemen2012}). This spatial resolution is lower than those in other simulations of CME eruptions \citep{Aulanier2010,Jiang2021}, but it is a compromise for the sake of saving computation resources for solving the thermal-MHD equations.

\subsection{Initial Conditions}
The initial magnetic field (similar to that in OHM; \citeauthor{Aulanier2010} \citeyear{Aulanier2010}) is composed of two potential bipolar fields:
\begin{equation}\label{eq7}
\begin{gathered}
B_x(t=0)=\Sigma_{m=1}^4c_m(x-x_m)r_m^{-3} \\
B_y(t=0)=\Sigma_{m=1}^4c_m(y-y_m)r_m^{-3} \\
B_z(t=0)=\Sigma_{m=1}^4c_m(z-z_m)r_m^{-3} \\
r_m=\sqrt{(x-x_m)^2+(y-y_m)^2+(z-z_m)^2},
\end{gathered}
\end{equation}
where ($c_1=60, x_1=0.9, y_1=0.3, z_1=-1.1$), ($c_2=-60, x_2=-0.9, y_2=-0.3, z_2=-1.1$), ($c_3=45, x_3=9, y_3=3, z_3=-11$), and ($c_4=-45, x_4=-9, y_4=-3, z_4=-11$) in dimensionless unit. The first (last) two monopoles are close to (far from) each other and locate shallow (deep), producing a strong (weak) potential field that decays rapidly (slowly) with altitude within the physical domain. The first potential field mimics the background field in an active region while the second one mimics the background field of the Sun on a global scale. The second potential field is set to ensure that the Alfvén speed at the upper part of the domain is large enough and thus to avoid shocks caused by the driving motion. The maximum dimensionless field strength in the plane $z=0$ (which is named as \textit{box bottom surface} in the following) is about 45. Such a strength is smaller than that in active regions in observations, while this is also a compromise for saving computation resources. 

The initial atmosphere is set to a hydrostatic corona with a uniform dimensionless temperature of 1. The box bottom surface corresponds to 3 Mm above the solar surface, and the initial dimensionless mass density at the box bottom surface is set to 1. The initial distribution of thermal pressure and mass density with altitude follows:
\begin{equation}\label{eq8}
\begin{gathered}
\nabla p=\rho g \\
p=\rho T \\
g=g_0\frac{R_{sun}^2}{(R_{sun}+0.3+z)^2},
\end{gathered}
\end{equation}
where $R_{sun}=69.55$ is the dimensionless solar radius and $g_0=-0.126$ is the dimensionless gravity acceleration at the solar surface. With such an initial atmosphere, the plasma $\beta$ along the vertical slit through the origin is less than 1 in the region $z<3.7$ (in dimensionless unit) at $t=0$. While, as the magnetic field evolves, the plasma $\beta$ further decreases: the plasma $\beta$ along the vertical slit through the origin is less than 1 in the region $z<4.8$ at $t=18$, in the region $z<6.0$ at $t=58$, and in the region $z<7.5$ at $t=68$, which ensures that the modelled CME (progenitor) is always located in a low plasma $\beta$ region which mimics the corona.

The initial velocity is set to zero and the parameter $\psi$ is also set to zero in the whole physical domain.

\subsection{Driving Motions and Line-tied Conditions}
In this work, we impose line-tied motions on the bottom boundary to drive the magnetic field and set special bottom boundary conditions to ensure the line-tied condition. Here, the line-tied condition means that the footpoint of the field line can only move horizontally according to the prescribed motion in the case of ideal MHD \citep{Aulanier2005}.

Inspired by the characteristics of the photospheric magnetic field evolution in observations \citep{Yang2004,Green2009,Schrijver2011}, two driving motions are considered in this simulation: a shearing motion that drives the initial potential field to a highly sheared state, and a converging motion that facilitates the flux cancellation near the polarity inversion line (PIL). The ``Simulation De'' is composed of two phases: the shearing phase ($0\le t\le18$) with only shearing motion applied and the converging phase ($18< t\le68$) with only converging motion applied. Such a setup differs from that in the previous work \citep{Zuccarello2015}, as here the converging motion is never switched off. We note that imposing only one type of motion at each phase makes the simulation easier to control and analyze: the shearing phase can be understood as a preparatory stage of the simulation to create a sheared magnetic field for the converging phase, during the latter of which we analyze the formation and eruption of the flux rope; while the converging phase is designed to fulfill the flux cancellation model.

\begin{figure}
\centering
\includegraphics[width=\hsize]{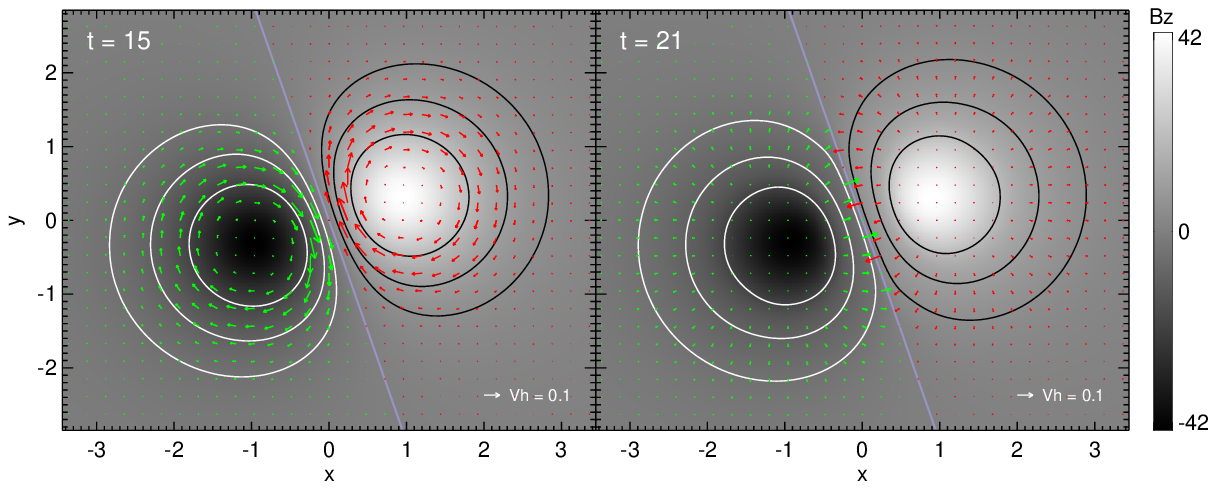}
\caption{\footnotesize{\textbf{Distribution of $B_z$ and imposed horizontal driving motions on the cell-center bottom surface.} The black (white) contours in the positive (negative) polarity are the contours of $B_z$, and the purple ones show the PILs. The red (green) arrows in the positive (negative) polarity in the left panel show the shearing flow at $t=15$, and those in the right panel show the converging flow at $t=21$. The arrow orientation denotes the direction of the motion, and its length represents the magnitude of the speed ($v_h$) . The scale of $v_h$ is shown at the bottom right corner of each panel. All parameters are in dimensionless units.}}
\label{figs1}
\end{figure}

The motions are imposed at the \textit{cell-center bottom surface} which is a horizontal surface at the altitude of the cell center of the first layer of the physical domain (cell layers in physical domain indexed by $k=1, 2, 3, ..., 96$ from the bottom to the top) at each time step to directly control the evolution of the magnetic field in the layer $k=1$. For ``Simulation De'', we set:

the shearing motion ($v_x^s, v_y^s, v_z^s$):
\begin{equation}\label{eq9}
\begin{gathered}
v_x^s(k=1; t)=\gamma(t)V_x^s(t) \\
v_y^s(k=1; t)=\gamma(t)V_y^s(t) \\
v_z^s(k=1; t)=0 \\
V_x^s(t)=v_0^{max}\Psi_0(t)\partial_y\Psi(t) \\
V_y^s(t)=-v_0^{max}\Psi_0(t)\partial_x\Psi(t) \\
\gamma(t)=\left\{
\begin{aligned}
\frac{1}{2}\tanh[3.75(t-1)]+\frac{1}{2} & & & & & & 0\le t<16 \\
-\frac{1}{2}\tanh[3.75(t-17)]+\frac{1}{2} & & & & & & 16\le t\le18, 
\end{aligned}
\right.
\end{gathered}
\end{equation}

and the converging motion ($v_x^c, v_y^c, v_z^c$):
\begin{equation}\label{eq10}
\begin{gathered}
v_x^c(k=1; t)=\gamma(t)V_x^c(t) \\
v_y^c(k=1; t)=\gamma(t)V_y^c(t) \\
v_z^c(k=1; t)=0 \\
V_x^c(t)=v_0^{max}\Psi_0(t)\partial_x\Psi(t) \\
V_y^c(t)=v_0^{max}\Psi_0(t)\partial_y\Psi(t) \\
\gamma(t)=\frac{1}{2}\tanh[3.75(t-19)]+\frac{1}{2} \ \ \ \ \ \ 18<t\le68,
\end{gathered}
\end{equation}
where
\begin{equation}\label{eq11}
\Psi(t)=\exp[-\Psi_1(\frac{B_z(k=1; t)}{B_z^{max}(k=1; t)})^2],
\end{equation}
and $\Psi_1=5.5$ for the shearing motion and $\Psi_1=27.5$ for the converging motion. The maximum speed of the driving motion $v_0^{max}$ is set to 0.16 (in dimensionless unit) for both the shearing and the converging motions. The function $\Psi_0(t)$ is used to keep the maximum of the term $\sqrt{V_x^2+V_y^2}$ always $v_0^{max}$. As shown in Fig. \ref{figs1}, the shearing motion is along the tangent direction of the contour of $B_z(k=1)$, while the converging motion is perpendicular to that tangent direction. Near the PIL, the shearing motion is antiparallel on two sides of the PIL, while the converging motion is towards the PIL.

It should be noted that the maximum speed of the driving motions is set to roughly 20 times of that in observations to speed up the simulation. While, it is still acceptable as it is less than the Alfvén speed and the sound speed in the core-area affected by the driving motion: at $t=0$, in the domain that $-4.20\le x\le4.20$, $-4.20\le y\le4.20$, and $0\le z\le8.36$ (in dimensionless unit), the ratio of $v_0^{max}$ to the average Alfvén speed is 0.101 and the ratio of $v_0^{max}$ to the maximum Alfvén speed is 0.004; the ratio of $v_0^{max}$ to the sound speed is constant of 0.124 in this domain. 

Several settings are adopted to achieve the line-tied condition at the cell-center bottom surface. First, the GLM parameter $\psi$ is fixed to zero in the layer $k=1$, and we set a symmetric boundary condition centered at the cell-center bottom surface for $\psi$ (see Section \ref{bc} for more details), so all components of the term $\nabla\cdot(\psi\boldsymbol{I})$ is zero in the layer $k=1$. Second, for ``Simulation De'', in the layer $k=1$, the dissipation term in the Equation \ref{eq3} is modified into:
\begin{equation}\label{eq12}
\frac{\partial \boldsymbol{B}}{\partial t}+\nabla\cdot(\boldsymbol{v}\boldsymbol{B}-\boldsymbol{B}\boldsymbol{v}+\psi\boldsymbol{I}) = \left\{
\begin{aligned}
0 & & & & & 0\le t\le18 \\
\eta(\frac{\partial^2}{\partial x^2}+\frac{\partial^2}{\partial y^2})\boldsymbol{B} & & & & & 18<t\le68.
\end{aligned}
\right.
\end{equation}
During the shearing phase of ``Simulation De'' ($0\le t\le18$), the dissipation term in this layer equals zero so that it has no effect on the magnetic field. The resistivity $\eta$ is fixed to zero in the layer $k=1$ and set to be uniform ($\eta=10^{-4}$; in dimensionless unit) in the whole region (including both the physical domain and the ghost cell layers) except the layer $k=1$ during the shearing phase. However, during the converging phase of ``Simulation De'' ($18< t\le68$), we adopt a two-dimensional (2D) dissipation term \citep{Aulanier2010} in the layer $k=1$ to allow the flux cancellation close to the PIL. A larger, uniform resistivity ($\eta=4\times10^{-4}$; in dimensionless unit), which facilitates the flux cancellation, is set in the whole region including the layer $k=1$.

In addition, the shearing motion in Equation \ref{eq9} satisfies that the z-component of the term $\nabla\cdot(\boldsymbol{v}\boldsymbol{B}-\boldsymbol{B}\boldsymbol{v})$ is zero, so analytically $B_z$ in the cell-center bottom surface should remain invariable during the shearing phase. To achieve this feature more accurately, we enforce that $B_z$ remains invariable in the layer $k=1$ during this phase.

\subsection{Boundary Conditions}\label{bc}
The mp5-limited reconstruction method uses three ghost cell layers at each boundary. The mesh in the ghost cell layer is also stretched in the same way as that in the physical domain. In all ghost cell layers, the background magnetic field $\boldsymbol{B_0}$ remains invariable; the two components of the magnetic field $\boldsymbol{B_1}$ which are parallel to the boundary surface are derived by one-sided third-order equal-gradient extrapolation, and the component of $\boldsymbol{B_1}$ which is normal to the boundary surface is derived under the constraint of the $\nabla\cdot\boldsymbol{B}=0$ condition. Such a setting will keep the divergence of the magnetic field zero in the layer $k=1$, so it is reasonable to set $\psi$ to zero in this layer.

We assume that at each time step, the atmosphere in both the bottom ghost cell layer and the layer $k=1$ is hydrostatic. Thus, the thermal pressure in the bottom ghost cell layer can be derived with centered difference. We further assume that the dimensionless temperature in the ghost cell layer is constant of 1, same as the initial temperature, and then the mass density is derived by the ideal gas law. The thermal pressure and the mass density in the top ghost cell layers are derived with the same method. In the four side ghost cell layers, the thermal pressure and the mass density are fixed to their initial values.

The velocity in the bottom ghost cell layer $k=n$ (layer indexed by $k=-1, -2, -3$ from the top to the bottom of the bottom ghost cells) is derived from the velocity in the layer $k=1$ and that in the layer $k=1-n$ by the linear extrapolation. This boundary condition is the same as the antisymmetric boundary condition for $v_z$ and the pseudo antisymmetric boundary condition for $v_x$ and $v_y$ at the bottom boundary in the previous work \citep{Aulanier2005}. The velocity in top ghost cell layers is derived by the one-sided second-order zero-gradient extrapolation, and the z-component is forced to be no less than zero to avoid possible downward flows. The velocity in the side ghost cell layers is set by the antisymmetric boundary condition centered on the side surface of the box.

To achieve the goal of removing the contribution of $\partial_z\psi$ in the layer $k=1$, a symmetric bottom boundary condition for $\psi$ is set centered on the cell center of this layer. The aim is fulfilled more precisely by replicating the flux of $\psi$ from the top surface of the layer $k=1$ to its bottom surface (box bottom surface). In the top and the side ghost cell layers, $\psi$ is fixed to zero.

\subsection{Overview of Modelled CME Event}
\begin{figure}
\centering
\includegraphics[width=0.8\hsize]{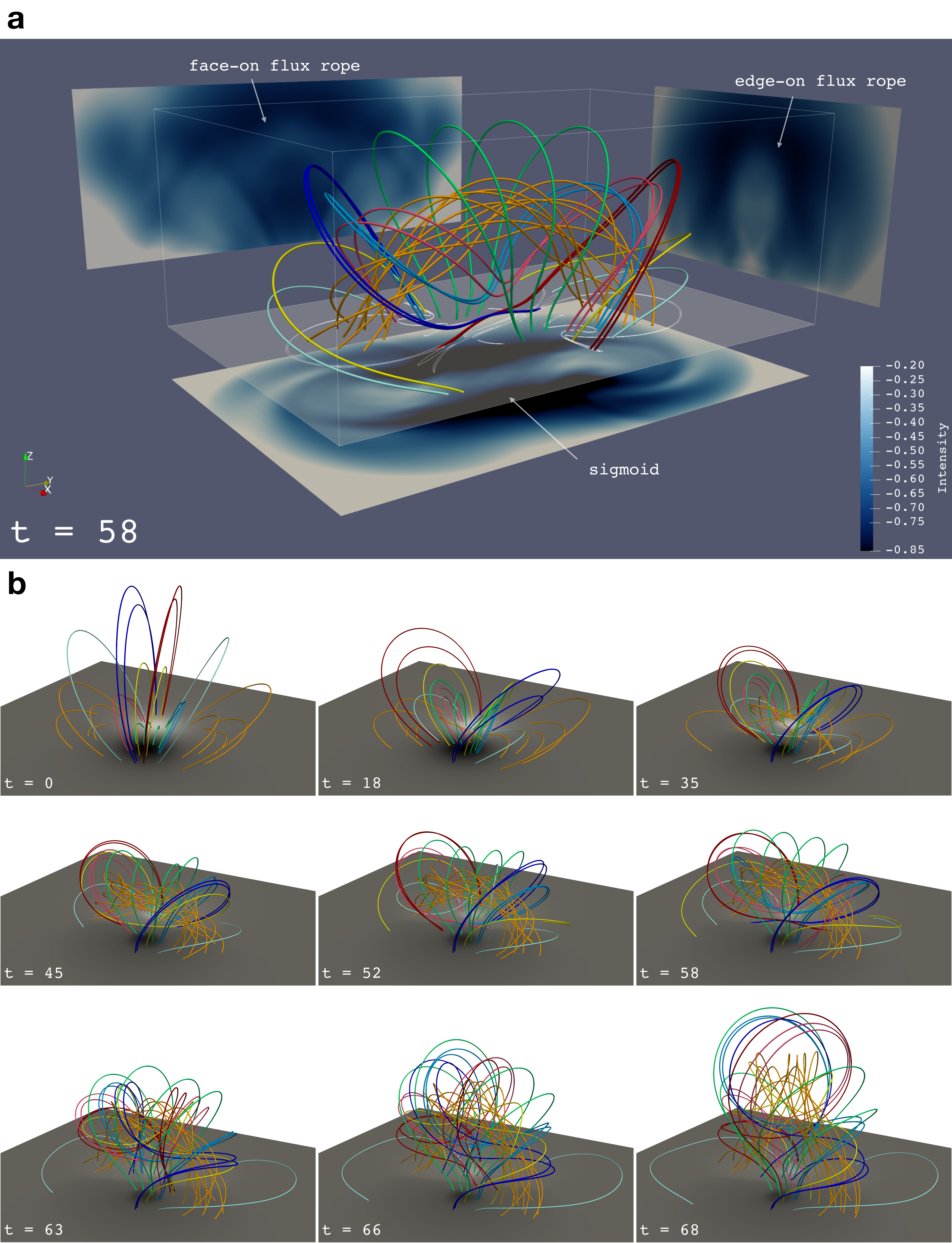}
\caption{\footnotesize{\textbf{Overview of the formation and eruption of the modelled flux rope.} (a) Field lines and percentage difference of synthetic EUV images at the AIA 335 \AA\ at $t=58$. The left, right, bottom images are the synthetic images observed from $x$, $y$, and $z$ directions, respectively. The translucent surface corresponds to cell-center bottom surface and the box is used for integrating the emissivity. The white contours on the bottom surface represent those of QSL footprints. The three side views of this panel are shown in Fig. \ref{figs9}c. (b) Evolution of magnetic field lines. The sub-panel at $t=58$ is the same as panel a. The cyan field line at $t=35$ and the yellow field line at $t=45$ are tangent to the bottom surface at their dips, representing the field lines traced from the BP. The light blue and light red field lines at $t=52$ and the dark blue and dark red field lines at $t=58$ exhibit those passing through the HFT. The cell-center bottom surface shows the distribution of $B_z$. White (black) represents the positive (negative) polarity. The scale of $B_z$ is the same in each panel. An animation of panel b is available, showing the evolution of field lines from $t=0$ to $t=68$ in the simulation. The duration of the animation is 23 s.}}
\label{fig1}
\end{figure}

The evolution of the modelled CME progenitor and CME in ``Simulation De'' is shown in Fig. \ref{fig1}. During the shearing phase ($0\le t\le18$), the initial potential field is driven to a highly sheared state under the shearing motion (Fig. \ref{fig1}b). After that, during the converging phase ($18<t\le68$), a pre-eruptive flux rope composed of twisted field lines is first formed by magnetic reconnection under the converging motion, and then it erupts as a CME (Fig. \ref{fig1}b). The pre-eruptive flux rope is hot and visible in synthetic EUV images mimicking AIA 335 $\textup{\AA}$ observations (Fig. \ref{fig1}a and Fig. \ref{figs9}).

\begin{figure}
\centering
\includegraphics[width=\hsize]{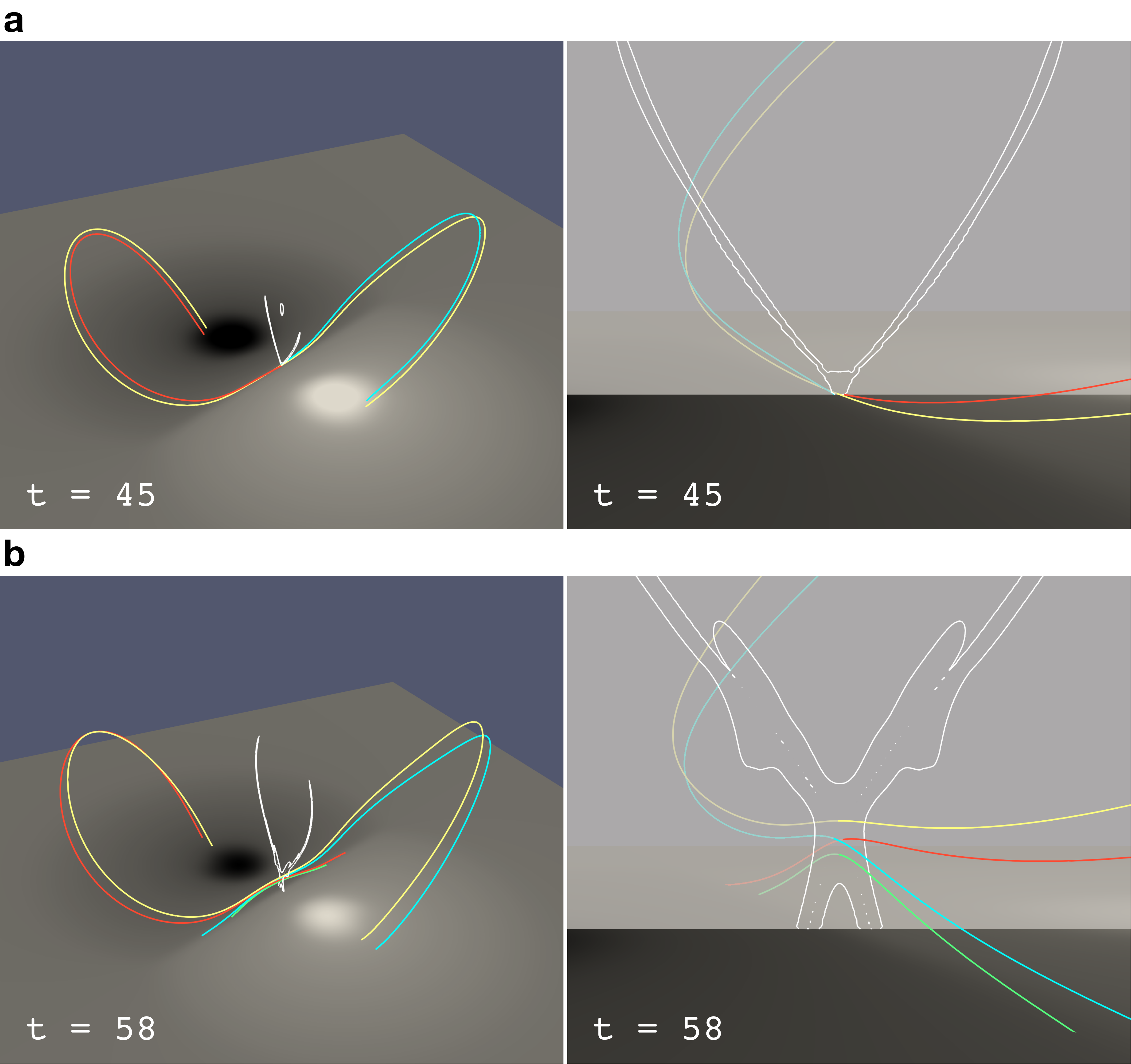}
\caption{\footnotesize{\textbf{Sketches of reconnections.} (a) A sketch of the BP reconnection. The BP reconnection occurs between two sheared arcades (symbolized by the red and the blue field lines) at two sides of the BP and forms a flux rope field line (symbolized by the yellow field line) which is tangent to the bottom surface at its dip. The cell-center bottom surface shows the distribution of $B_z$. The white contours show the profiles of the QSLs in the plane $y=0$. The white vertical plane in the right sub-panel is the plane $y=0$. (b) A sketch of the HFT reconnection. The HFT reconnection occurs between two sheared arcades (symbolized by the red and the blue field lines) at right and left sides of the HFT. It forms a flux rope field line (symbolized by the yellow field line) whose dipped part is detached from the bottom surface and at the upper part of the HFT, and a low-lying loop (symbolized by the green field line) at the lower part of the HFT. The cell-center bottom surface, white contours, and the white vertical plane have the same meanings as those in panel a.}}
\label{figs2}
\end{figure}

The initial magnetic topology of the reconnection region is a bald patch (BP; \citeauthor{Titov1993} \citeyear{Titov1993}; during $20\le t\le45$), characterized by $(\textbf{\textit{B}}\cdot\nabla)B_z>0$ where $B_z=0$. As shown by the quasi-separatrix layers (QSLs; \citeauthor{Priest1995} \citeyear{Priest1995}), such a BP topology later gradually bifurcates into a hyperbolic flux tube (HFT; \citeauthor{Titov2002} \citeyear{Titov2002}; since $t=46$; see Fig. \ref{fig3} and Fig. \ref{fig4}f). An obvious difference between the BP reconnection and the HFT reconnection is that the former forms flux rope field lines tangent to the bottom surface at their dips (Fig. \ref{figs2}a), while the latter forms flux rope field lines whose dipped parts are detached from the bottom surface and low-lying loops (Fig. \ref{figs2}b).

\section{Kinematics of CME Progenitor and CME}
The kinematics of CME progenitor and CME in ``Simulation De'' is estimated by measuring the height of the apex of an overlying field line right above the flux rope. Such an estimation method in simulations is consistent with the estimation method in observations where the kinematics of the hot channel is measured at its leading edge. The overlying field line is selected as one traced from a same fixed point at the center of the positive polarity and on the cell-center bottom surface. The velocity at this fixed point is extremely small, very close to or equal to zero, in all phases of ``Simulation De'', which ensures that the footpoint of this field line hardly moves and its height evolution can well reflect that of the CME (progenitor).

\begin{figure}
\centering
\includegraphics[width=\hsize]{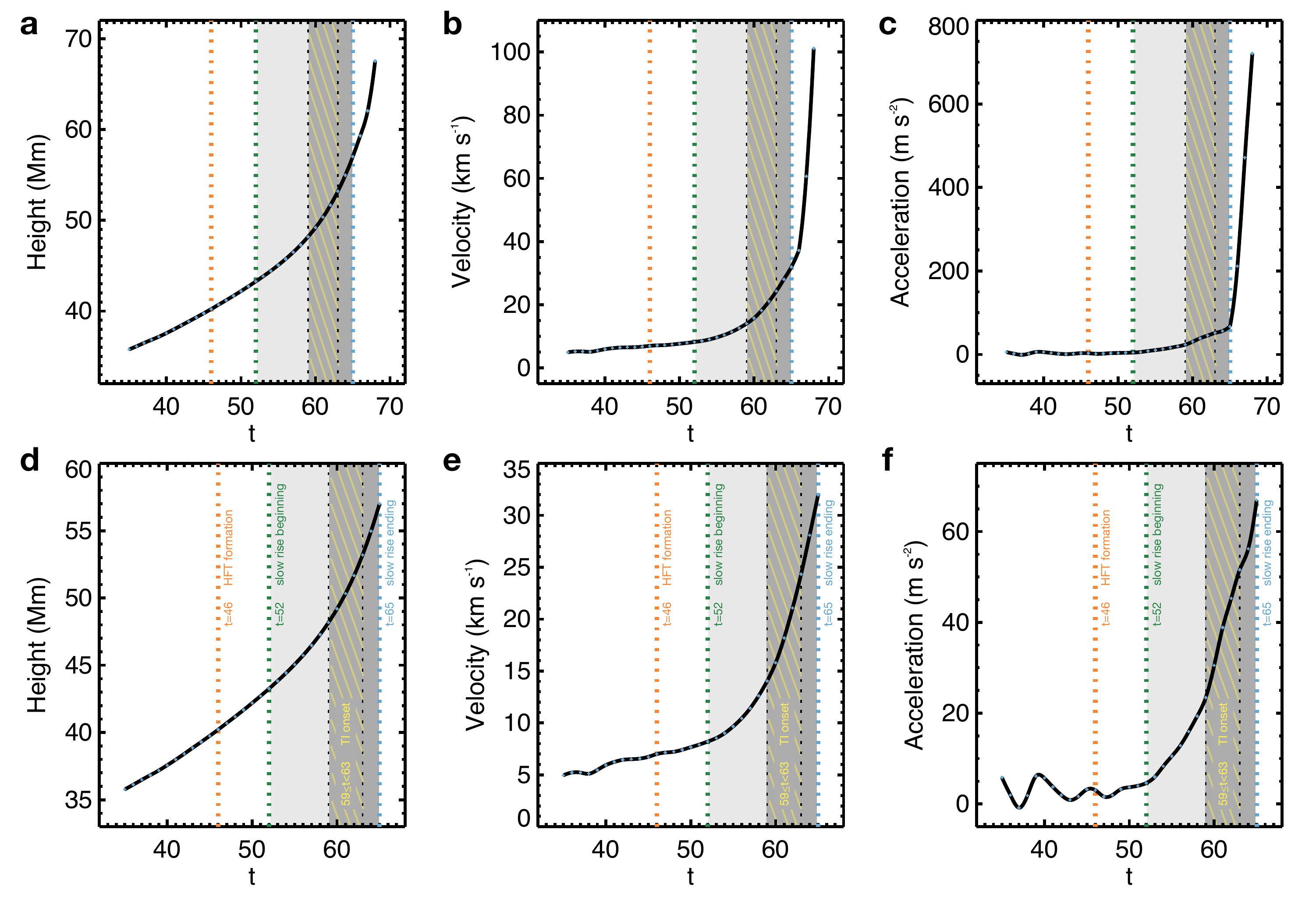}
\caption{\footnotesize{\textbf{Kinematics of the CME progenitor and CME.} (a)-(c) Kinematics of the CME progenitor and CME during $35\le t\le68$. In each sub-panel, the blue points show the measured or derived data and the black curve shows the fitting curve of data with the cubic spline interpolation method. The orange dashed line represents the time when the HFT first appears. The green and blue dashed lines mark the beginning and the end of the slow rise phase, respectively. The light grey and dark grey regions show the earlier stage and the later stage of the slow rise phase, respectively. The two black dashed lines and the yellow oblique-line-region between them mark the time range of the torus instability onset. (d)-(f) Same as panels a-c but for the kinematics during $35\le t\le65$.}}
\label{fig2}
\end{figure}

The kinematic of CME progenitor and CME, as shown in Fig. \ref{fig2}, is composed of three phases which are comparable to the three phases in observations, even though the simulation is merely designed for CME progenitor formation and CME eruption and certainly not to fit the kinematics a priori. The CME progenitor first rises quasi-statically with a tiny and fluctuating acceleration (during $35\le t<52$; Fig. \ref{fig2}f), consistent with the behavior in the quasi-static phase in observations \citep{Xing2018}. As a result, the velocity-time curve is in a quasi-linear form in this phase (Fig. \ref{fig2}e). We note that the decaying fluctuation of the acceleration may reflect the process of the CME progenitor gradually approaching the quasi-equilibrium state under the driving motion. Later (during $52\le t\le65$), the acceleration of the flux rope continuously increases to tens of m s$^{-2}$ (Fig. \ref{fig2}f) and thus its velocity increases quickly (Fig. \ref{fig2}e). This phase corresponds to the slow rise phase in observations and thus marks the initiation process of the modelled CME, as this phase is just before the impulsive acceleration of the CME and the evolution of acceleration in it is significantly different from those in phases both before and after it. Even, the acceleration in this phase is comparable to that in the slow rise phase in observations (also tens of m s$^{-2}$; \citeauthor{Cheng2020} \citeyear{Cheng2020}). The last phase ($65< t\le68$) is considered to correspond to the impulsive acceleration phase in observations, as the modelled CME acceleration impulsively rises to hundreds of m s$^{-2}$ in this period (Fig. \ref{fig2}c).

The eruption of CME is most likely triggered by the torus instability (TI), which occurs in the later stage of the slow rise phase (at a moment during $59\le t<63$; see the analysis of torus instability in Appendix \ref{AppA}). Other potential eruption-trigger mechanisms are ruled out: (1) The HFT reconnection can not trigger the eruption as the flux rope fails to erupt in the presence of the HFT in the control simulation ``Simulation Ne'' (see Appendix \ref{AppA}). (2) The long and thin current sheet, which is usually necessary for the fast magnetic reconnection \citep{Jiang2021}, does not show up before $t=65$ in this simulation. Therefore, the fast magnetic reconnection, even if it sets in this simulation, occurs later than the onset of the torus instability and does not play a key role in triggering the eruption.

Lastly, it should be noted that the behavior of the acceleration of the CME progenitor is less affected by the absolute speed of the driving motion during the quasi-static phase in this simulation. The reason is that the rising of the CME progenitor is a response to the driving motion in this phase, while the maximum speed of driving motions is a constant. Therefore, even if the absolute speed of the driving motion is large, the acceleration of the CME progenitor, as the first order derivative of the velocity with respect to time, is still almost close to 0. Therefore, the quasi-static phase in our simulation can be compared to that in observations.

\section{Mechanisms of Kinematics}
\subsection{Triggering Mechanism of Slow Rise Phase}
The transition from the quasi-static phase to the slow rise phase occurs (at $t=52$) just after the first appearance of HFT (at $t=46$). This indicates that the HFT reconnection is likely the mechanism triggering the slow rise phase.

This speculation is supported by detailed analyses on the relation between the reconnection and the kinematics. Fig. \ref{fig3}a,f reveal that the upward outflow of the reconnection is very slow at the stage of BP (e.g., $t=45$) and the early stage of HFT ($46\le t<52$), but it is markedly accelerated as the HFT is rapidly built up (e.g., from $t=52$ to $t=58$). Quantitatively, the growth rate of the reconnection upward outflow velocity is temporally consistent with the acceleration of the CME progenitor, and both of them start to increase shortly after the first appearance of HFT (Fig. \ref{fig3}g). We consider that the velocity of the reconnection upward outflow reflects the ability of reconnection to contribute to the flux rope. Therefore, the above results indicate that the slow rise is triggered when the contribution from reconnection starts to quickly increase after the HFT forms. In other words, the slow rise is triggered by the HFT reconnection. We eliminate the possibility that the torus instability triggers the slow rise, since the failed eruption in the ``Simulation Ne'' implies that the torus instability, as the triggering mechanism of the eruption, does not occur before $t=58$ in the ``Simulation De'' (see Appendix \ref{AppA}).

\subsection{Driving Mechanisms of Slow Rise Phase}
We further investigate the mechanisms driving the slow rise of the flux rope. Considering that the torus instability occurs during the slow rise phase, we divide the slow rise phase into two stages, i.e., the earlier stage ($52\le t\le58$) and the later stage ($59\le t\le65$), and analyze the driving mechanisms in these two stages separately.

\begin{figure}
\centering
\includegraphics[width=\hsize]{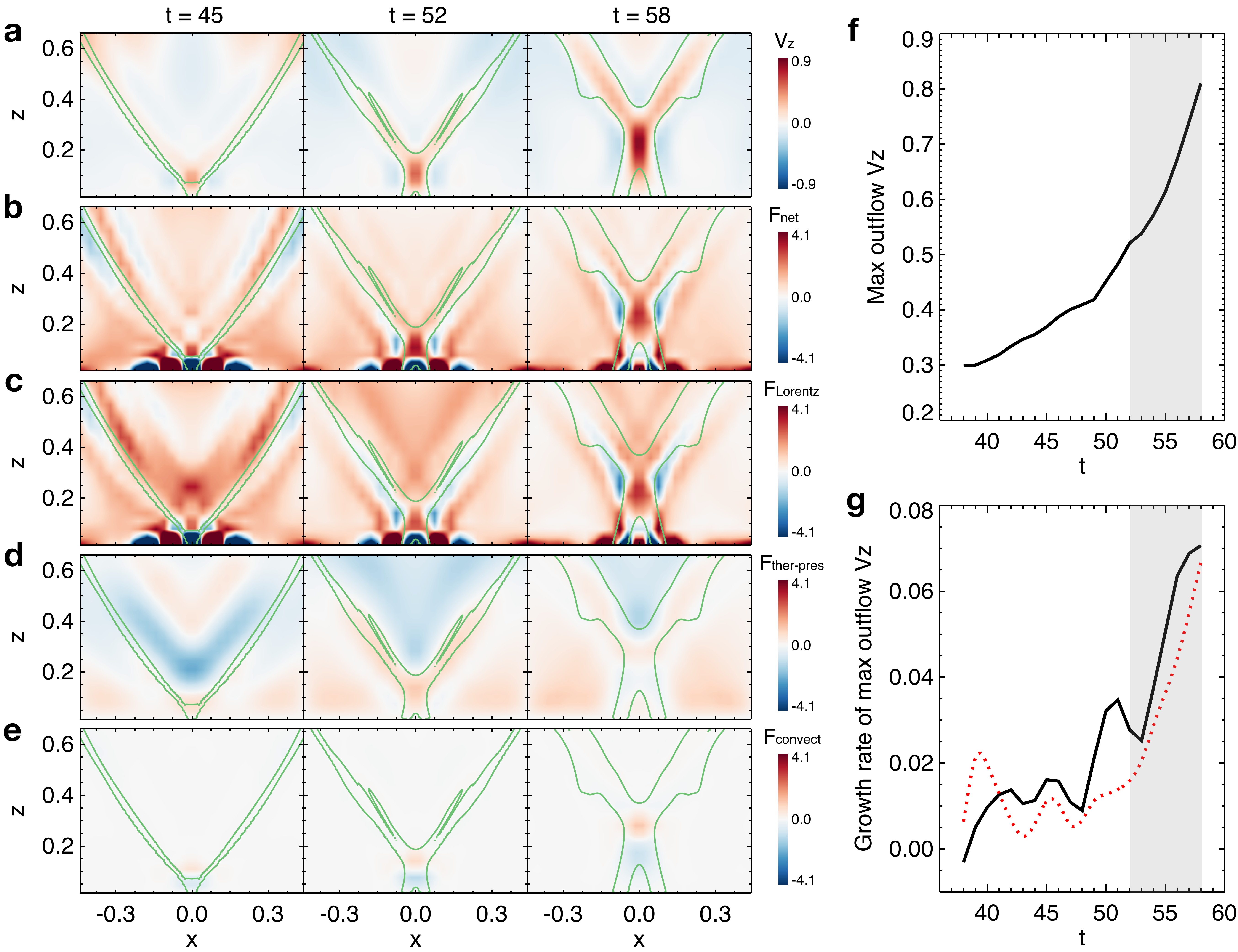}
\caption{\footnotesize{\textbf{Evolutions of forces and velocity in the reconnection region.} (a)-(e) Component of forces and velocity in the z-direction in the plane $y=0$ at $t=45$, $t=52$, and $t=58$. They are in order as follows: velocity ($v_z$; panel a), net force ($F_{net}$; panel b), Lorentz force ($F_{Lorentz}$; panel c), thermal pressure gradient force ($F_{ther-pres}$; panel d), and force related to the convection ($F_{convect}$; panel e). $F_{net}$ is the sum of the z-components of Lorentz force, thermal pressure gradient force, gravity, and viscous force. All of forces and velocity are in dimensionless units. The green contours outline the QSLs, indicating that the magnetic topology of the reconnection region is a BP at $t=45$ and an HFT at $t=52$ and $t=58$. (f) The maximum z-component of velocity ($v_z$; in dimensionless unit) of the reconnection outflow along the vertical slit through the origin during $38\le t\le58$. The grey region shows the slow rise phase before $t=58$. (g) The black curve shows the growth rate of the maximum $v_z$ of the reconnection outflow during $38\le t\le58$, which is derived by taking the derivative of the maximum $v_z$ with respect to time. The red dashed curve represents the fitting curve of the flux rope acceleration before $t=58$, which is the same as that in Fig. \ref{fig2}f but numerically scaled down by a factor of 290. The grey region has the same meaning as that in panel f.}}
\label{fig3}
\end{figure}

For the earlier stage, the motion of the flux rope is controlled by a net upward force on its lower part and a net downward force on its upper part (see Fig. \ref{figs5}b, for an example at $t=58$). The Lorentz force and the thermal pressure gradient force are the major contributors to the net force, while the gravity and the viscous force are of little importance (Fig. \ref{figs5}). Especially, the maxima of the Lorentz force, the net force, and the velocity in the flux rope region are all concentrated near the HFT (Fig. \ref{figs5}a,b) during this stage. This implies that it is the HFT reconnection that powers the flux rope rise in the earlier stage of the slow rise phase. In addition, the synchronized evolutions of the flux rope acceleration and the growth rate of the reconnection upward outflow velocity in Fig. \ref{fig3}g also indicate that the slow rise is driven by the HFT reconnection during the earlier stage.

The HFT reconnection drives the slow rise in the earlier stage, on the one hand, by adding twisted concave-upward field lines to the lower part of the flux rope (Fig. \ref{figs6}). These field lines provide magnetic tension to the flux rope (Fig. \ref{figs5}a; known as the slingshot effect), enhance the force imbalance in the flux rope, and finally accelerate the flux rope. On the other hand, the reconnection upward outflow in the HFT is convected upwardly into the flux rope (Fig. \ref{fig3}a,e), and thus it can directly promote the flux rope rise. Specifically, the plasma in the reconnection upward outflow is accelerated by a net upward force which is dominated by the Lorentz force and the thermal pressure gradient force (Fig. \ref{fig3}). The upward Lorentz force consists of magnetic tension force mainly at the HFT top and magnetic pressure gradient force mainly at the HFT center (Fig. \ref{figs7}), the former of which is provided by the newly formed concave-upward field lines and the latter of which is associated to the accumulation of the sheared/guide component of the magnetic field. The upward thermal pressure gradient force is due to an increase in the gas pressure at the flanks of the flux rope bottom, which is a result of the density pileup pinched by the reconnection inflow (Fig. \ref{figs7}).

Furthermore, since the HFT appears, the two footprints of the HFT continuously separate from each other in the direction perpendicular to the PIL, and meanwhile they extend in the direction parallel to the PIL to both sides until they are completely separated (Fig. \ref{fig4}a-c; shown by the QSL footprint, which manifests as the flare ribbon in observations during the eruption \citep{Zhao2016}, but probably invisible before the eruption due to the weak reconnection rate). The reconnection electric field within the HFT, which represents the reconnection rate \citep{Forbes1984}, also shows a gradually increasing trend since $t=46$ and until $t=59$ (Fig. \ref{fig4}h; also see Appendix \ref{AppC} for more details). These results strongly suggest that the HFT and its associated reconnection are gradually built up during the earlier stage of the slow rise phase. Both the slingshot effect and the outflow effect of the HFT reconnection thus become stronger, which leads to the continuous increase of acceleration of the CME progenitor in this stage.

For the later stage of the slow rise phase, we argue that the HFT reconnection still has a considerable contribution to the rising motion, since the HFT reconnection becomes even more stronger in this stage, shown by the quickly increasing reconnection electric field (see that during $59\le t\le65$ in Fig. \ref{fig4}h). Moreover, the torus instability also promotes the slow rise in the later stage since it starts. On the one hand, the hoop force could directly accelerate the flux rope \citep{Kliem2006}; on the other hand, the torus instability could indirectly contribute to the slow rise by enhancing the HFT reconnection, evidenced by that the HFT reconnection electric field immediately shows a quick increase once the torus instability sets in (Fig. \ref{fig4}h). This enhancement may be due to that the current structure in the HFT could become longer as the flux rope rises higher and faster with the contribution from the torus instability, which well reflects the close coupling of torus instability and HFT reconnection in the later stage.

\subsection{Triggering and Driving Mechanisms of Impulsive Acceleration Phase}
A fast magnetic reconnection likely occurs during the impulsive acceleration phase, as a thin and long current sheet, which is usually the necessary condition for the fast magnetic reconnection, is formed at the start of the impulsive acceleration phase ($t\sim66$) and soon built up since then (Fig. \ref{figs8}). The other evidence for the fast magnetic reconnection is the impulsively increasing reconnection electric field in the reconnection region during the impulsive acceleration phase (Fig. \ref{fig4}g). Therefore, we consider that the impulsive acceleration phase ($65<t\le68$) is likely triggered as the magnetic reconnection coupled to the torus instability transforms from the relatively weaker HFT reconnection to the relatively stronger fast magnetic reconnection, and that the phase is continuously driven by the close coupling of the fast magnetic reconnection and the torus instability which induces an impulsive acceleration.

\section{Heating of CME Progenitor}
Benefitting from the thermal-MHD simulation, we also study the thermal properties of the CME progenitor. The modelled CME progenitor is significantly heated and visible in EUV synthetic images at the AIA 335 $\textup{\AA}$ (Fig. \ref{fig1}a and Fig. \ref{figs9}; see Appendix \ref{AppD} for more details). It is channel-like in the face-on view, elliptical in the edge-on view, and sigmoidal in the top view. It is very similar to pre-eruptive hot channels discovered in observations \citep{Zhang2012,Patsourakos2013,Cheng2014b}, except that our modelled one is somewhat less hot than those in observations due to the weak magnetic field adopted in our simulation. The temperature of the modelled CME progenitor is expected to increase with a stronger magnetic field, as the former is roughly proportion to the square of the later. Regardless of the specific temperature, the modelled bright structure basically corresponds to the pre-eruptive flux rope (Fig. \ref{figs9}), which supports the view that the hot channel represents the hot flux rope in observations \citep{Zhang2012}.

In addition, taking the snapshot at $t=58$ as an example, an analysis of the heating sources reveals that the CME progenitor is mainly heated by the Ohmic dissipation (Fig. \ref{fig6}). Especially, one can find that at the bottom and side boundaries of the flux rope, the Ohmic dissipation is relatively stronger and its profile is highly consistent with that of the HFT and the QSLs (Fig. \ref{fig6}). This result indicates that the magnetic reconnection in the HFT and the QSLs surrounding the flux rope effectively heats the local plasma while it forms the pre-eruptive flux rope, which confirms the conjecture in the previous 3D CME model \citep{Janvier2014}. In addition, the bulk heating within the flux rope results from the setup of a uniform resistivity in the simulation domain.

\section{Summary: A Complete Route of CME Initiation}
Based on our simulation, we present a complete route of the CME initiation as summarized in Fig. \ref{fig5}. In short, during the CME initiation, the slow rise motion is first triggered and driven by the HFT reconnection and later driven by an early coupling of the HFT reconnection and the torus instability.

In detail, the kinematic evolution around the CME initiation is described as the following. With the magnetic topology of the reconnection region changing from the BP to the HFT, the HFT reconnection starts to promote the flux rope rising with its slingshot effect and reconnection upward outflow. As the HFT reconnection is developing, the effect of the HFT reconnection to the flux rope rise gradually increases and eventually starts the slow rise phase/CME initiation. Afterwards, at the earlier stage of the slow rise phase/CME initiation, the pre-eruptive flux rope rises slowly with an increasing acceleration owing to the progressively enhanced HFT reconnection. With the HFT-dominated slow rise going on, the flux rope later reaches a critical height where the torus instability sets in. The onset of the instability marks the onset of the CME eruption in the physical sense, i.e., the flux rope being out of equilibrium. Then, at the later stage of the slow rise phase/CME initiation, the torus instability induces a stronger HFT reconnection, and they together promote the slow rise of the flux rope. Finally, the impulsive acceleration of CME starts (the CME initiation ends) when the magnetic reconnection coupled to the torus instability transforms from the HFT reconnection to the fast magnetic reconnection, which is later than the eruption onset in the physical sense.

\section{Discussion}
In previous studies, the slow rise motion during the initiation process was explained by a single mechanism such as the eruption-trigger mechanism \citep{Zhang2001,Zhang2006} or the weak magnetic reconnection before the onset of eruption-trigger mechanisms \citep{Sterling2007,Savcheva2012,Cheng2020,Cheng2023}. In contrast, our work presented here reveals that the CME initiation is actually a multiple-physics coupled-process and its complete route is hardly to be explained by the previous single mechanism. Firstly, the HFT reconnection triggers the CME initiation and drives the CME progenitor to slowly rise in quasi-equilibrium. Secondly, after the torus instability triggers the eruption, the less energetic destabilization and the HFT reconnection jointly contribute to the slow rise/CME initiation for a while until the coupling of the destabilization and the fast magnetic reconnection begins and induces the impulsive acceleration of the CME.

For the HFT reconnection, our simulation especially provides insight into the question how it drives the flux rope during the initiation process. We find that both the Lorentz force and the thermal pressure gradient force play important roles in the slow rise phase, at least in its earlier stage. Quantitatively, for the flux rope field lines at the upper part of the HFT, the ratio of the maximum of the thermal pressure gradient force to that of the Lorentz force is roughly 40$\%$ at $t=52$ and 12$\%$ at $t=58$ (Fig. \ref{fig3}c,d; the plasma $\beta$ is roughly 0.01 at regions of the above-mentioned maximum of the force). This result for the first time confirms the previous conjecture on the key role of Lorentz force in causing the rise of pre-eruptive structures in the ``tether-cutting reconnection'' model \citep{Moore2001}. However, the contribution of thermal pressure gradient force to the slow rise is also not negligible, while this result could be further influenced by the plasma $\beta$ in the HFT reconnection region. This underscores the indispensability of a thermal-MHD simulation, beyond the zero-$\beta$ MHD simulations \citep{Aulanier2010,Zuccarello2015}, in the study of the CME initiation.

Another important finding in our work is that the impulsive acceleration of CMEs does not immediately starts at the onset of the torus instability but commences at a later moment when the coupling of the torus instability and the fast magnetic reconnection starts. Such a time delay may explain why the critical decay indices estimated in observations \citep{Zou2019,Cheng2020,Cheng2023} are sometimes larger than those derived theoretically \citep{Kliem2006,Demoulin2010}, as the observationally determined eruption onset time is probably later than the real onset time of the torus instability in these works. It should be mentioned that a larger magnetic field strength in observations compared to that in this simulation may lead to a larger kinematic acceleration, and thus the above-mentioned time delay may become shorter in observations. In addition, such a time delay indicates that the torus instability can only induce a relatively small acceleration at its beginning, compared to that in the impulsive acceleration phase. Lastly, we emphasize that the finding on the time delay benefits from the setup of our simulation, i.e., continuously imposing driving motion after the eruption is triggered. Such a setup is distinguished from that in previous works \citep{Aulanier2010,Zuccarello2015}, in latter of which the driving motion is switched off to relax the system during the eruption. Our simulation avoids the extra relaxation effect in previous works, which does not occur in observations but could also contribute to the time delay in simulations and pose a significant difficulty for analysis (as discussed in Appendix \ref{AppA2}).

Our simulation sheds light on the heating process of CME progenitors. However, it should be noted that the thermodynamics in our simulation is a simplification of that of the Sun for the sake of clear and concise results. First, the density and temperature profile of the atmosphere was simplified in the simulation by omitting the lower atmosphere (from the photosphere to the transition region) and setting a uniform coronal temperature. Second, some processes in flares, e.g., the radiative cooling and the chromospheric evaporation, were not taken into account in the simulation. Even so, we emphasize that these simplifications do not critically affect our result, i.e., the magnetic reconnection takes place in the current sheet surrounding the flux rope and builds up and heats the CME progenitor.

In summary, we successfully reproduce the complete early kinematics of a CME event with a 3D thermal-MHD simulation that is more physically realistic in this work. For the first time, we explicitly disentangle the complex CME initiation process by systematically investigating how various mechanisms play roles at different stages of the kinematics of a CME event.

\section*{acknowledgments}
We thank Yuhao Zhou, Jinhan Guo, Yang Guo, Rony Keppens, Can Wang, and Feng Chen for valuable discussions.
C.X. and G.A. acknowledge financial support from the French national space agency (CNES), as well as from the Programme National Soleil Terre (PNST) of the CNRS/INSU also co-funded by CNES and CEA.
C.X. also acknowledges financial support from China Scholarship Council.
X.C. and M.D.D. are supported by National Key R\&D Program of China under grants 2021YFA1600504 and NSFC under grant 12127901. 
The numerical calculations in this paper were done on the computing facilities in High Performance Computing Center of Nanjing University and the Cholesky computing cluster from the IDCS mesocentre at Ecole Polytechnique.
MPI-AMRVAC 2.2, the code used to perform the simulation in this paper is an open source code. It is available from the website \url{https://amrvac.org/index.html}. The code used to calculate the squashing degree is available from the website \url{https://github.com/el2718/FastQSL}.

\clearpage
\newpage

\clearpage
\newpage

\clearpage
\newpage

\clearpage
\newpage

\clearpage
\newpage

\clearpage
\newpage
\begin{figure}
\centering
\includegraphics[width=\hsize]{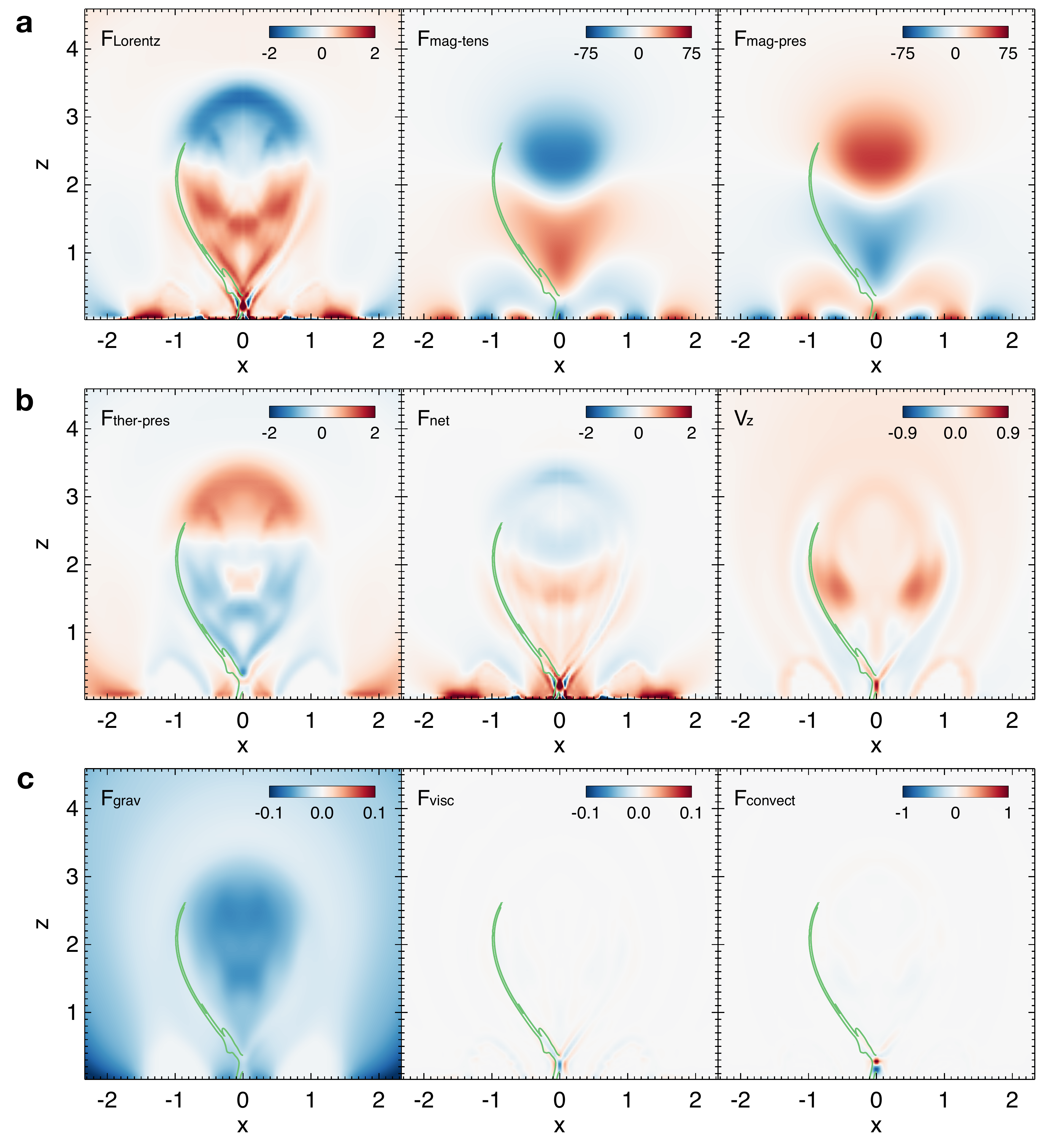}
\caption{\footnotesize{\textbf{Component of forces and velocity in the z-direction in the plane $y=0$ at $t=58$.} They are in order as follows: Lorentz force ($F_{Lorentz}$), magnetic tension force ($F_{mag-tens}$), magnetic pressure gradient force ($F_{mag-pres}$), thermal pressure gradient force ($F_{ther-pres}$), net force ($F_{net}$), velocity ($v_z$), gravity ($F_{grav}$), viscous force ($F_{visc}$), and force related to the convection ($F_{convect}$). $F_{net}$ is the sum of $F_{Lorentz}$, $F_{ther-pres}$, $F_{grav}$, and $F_{visc}$. Here the $F_{mag-tens}$ ($F_{mag-pres}$) is derived by first decomposing $\boldsymbol{B}\cdot\nabla\boldsymbol{B}$ ($-\nabla(B^2/2)$) to the normal direction of magnetic field and then further decomposing to the z-direction, given that the tangential component does not contribute to the acceleration. All of forces and velocity are in dimensionless units. The green contours represent the profiles of QSLs in the half panel of $x\le0$. See Appendix \ref{AppB} for more details for locating the flux rope boundary.}}
\label{figs5}
\end{figure}

\clearpage
\newpage
\begin{figure}
\centering
\includegraphics[width=\hsize]{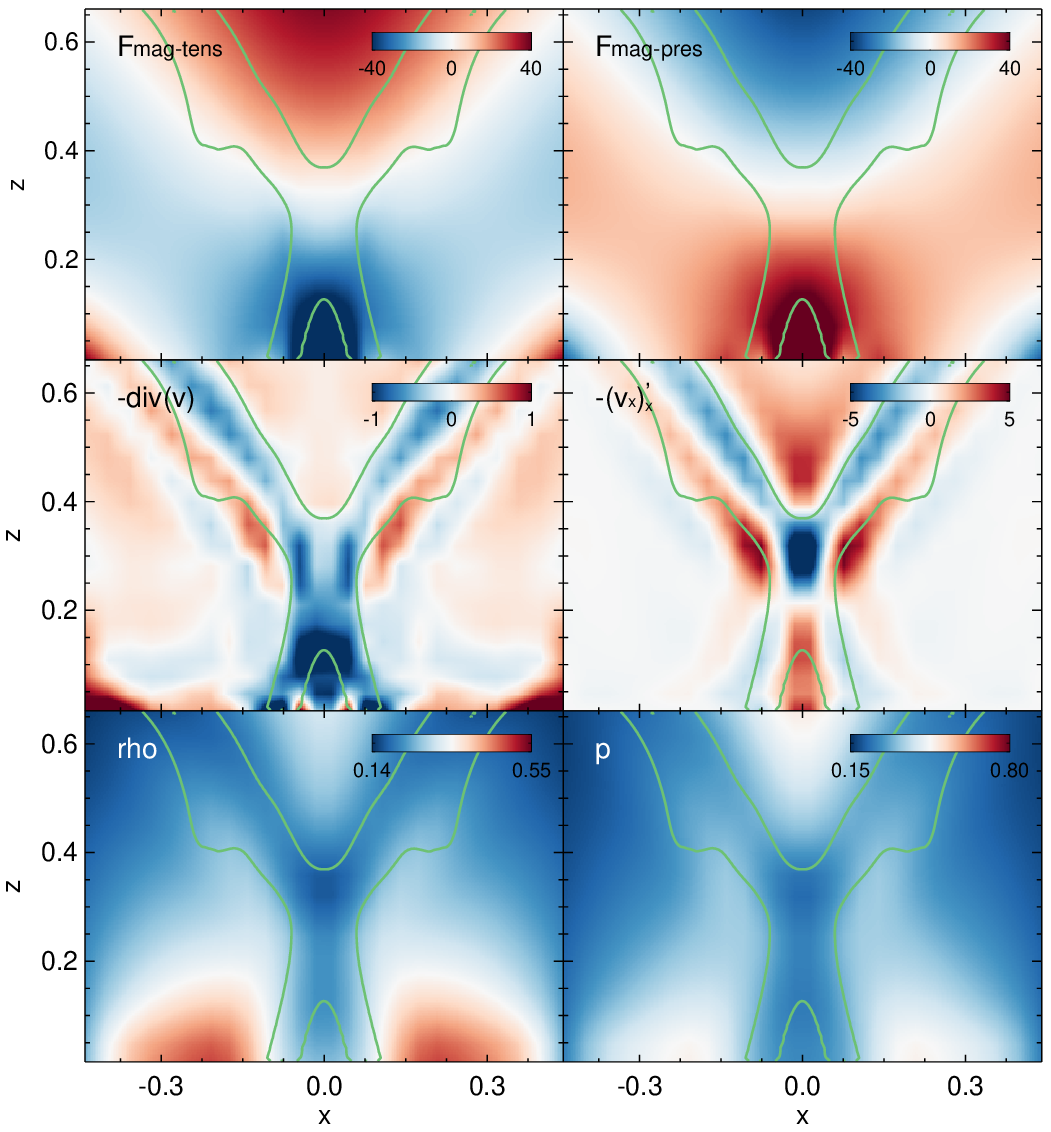}
\caption{\footnotesize{\textbf{Some parameters related to force analysis on the plane $y=0$ at $t=58$.} They are in order as follows: z-component of magnetic tension force ($F_{mag-tens}$), z-component of magnetic pressure gradient force ($F_{mag-pres}$), negative divergence of velocity ($-\textup{div}(v)$), negative partial derivative of x-component of velocity to $x$ ($-(v_x)_x^{'}$), mass density (rho), and thermal pressure ($p$). All parameters are in dimensionless units. The positive $-\textup{div}(v)$ and $-(v_x)_x^{'}$ at the two flanks of the flux rope bottom clearly show the compression caused by the reconnection inflow. The green curves represent contours of the QSLs.}}
\label{figs7}
\end{figure}

\clearpage
\newpage
\begin{figure}
\centering
\includegraphics[width=\hsize]{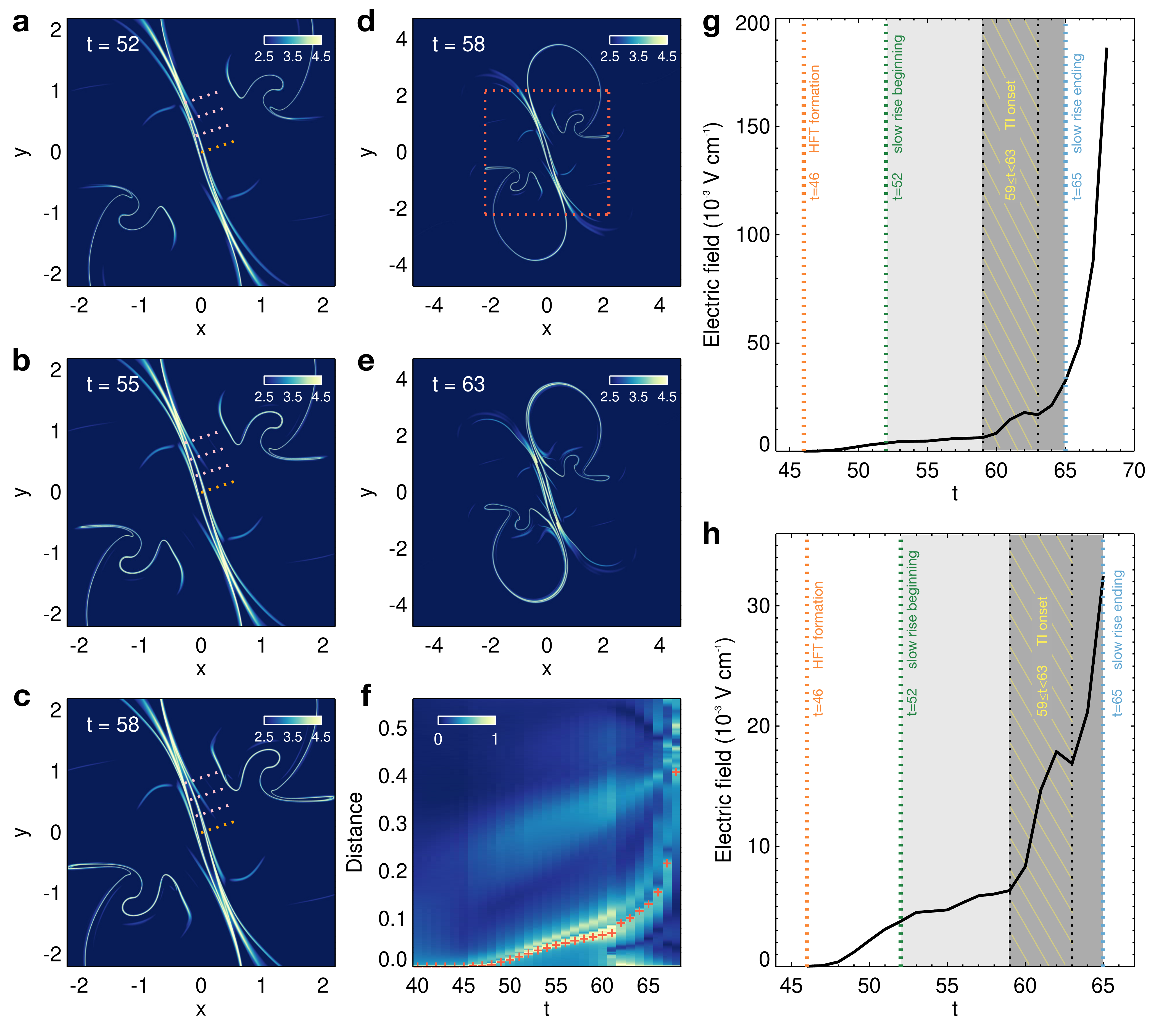}
\caption{\footnotesize{\textbf{Evolutions of HFT footprint and HFT reconnection electric field.} (a)-(c) Squashing degree $\textup{log}Q$ on the cell-center bottom surface at $t=52$, $t=55$, and $t=58$. The field of view is marked by the red dashed box in panel d. (d)-(e) Squashing degree $\textup{log}Q$ on the cell-center bottom surface at $t=58$ and $t=63$. The two hook-shape QSL footprints around $(X,Y)=(1,2)$ and $(X,Y)=(-1,-2)$ mark the boundaries of two footpoints of the flux rope, respectively. At the ends of straight parts of QSL footprints (around $(X,Y)=(-2,2)$ and $(X,Y)=(2,-2)$ in panel e), the curved QSL footprints demonstrate the rapid change in the connectivity of the field lines anchored there as a result of the driving motion. (f) Time-slice plot of the normalized $\textup{log}Q$. The slit is denoted by the orange dashed line in panels a-c. At each moment, $\textup{log}Q$ is normalized by the maximum $\textup{log}Q$ along the slit, and the red symbol marks the location of the local maximum of normalized $\textup{log}Q$ which corresponds to the HFT footprint. (g) Reconnection electric field $E$ during $46\le t\le68$. The reconnection electric field during $46\le t\le51$ is derived from the orange slit in panels a-c; the reconnection electric field during $52\le t\le68$ is averaged from the electric fields derived from four slits (represented by the orange and the pink dashed lines in panels a-c). The vertical dashed lines, light/dark grey regions, and yellow oblique-line-region have the same meanings as those in Fig. \ref{fig2}. (h) Same as panel g but for the reconnection electric field $E$ during $46\le t\le65$.}}
\label{fig4}
\end{figure}

\clearpage
\newpage
\begin{figure}
\centering
\includegraphics[width=\hsize]{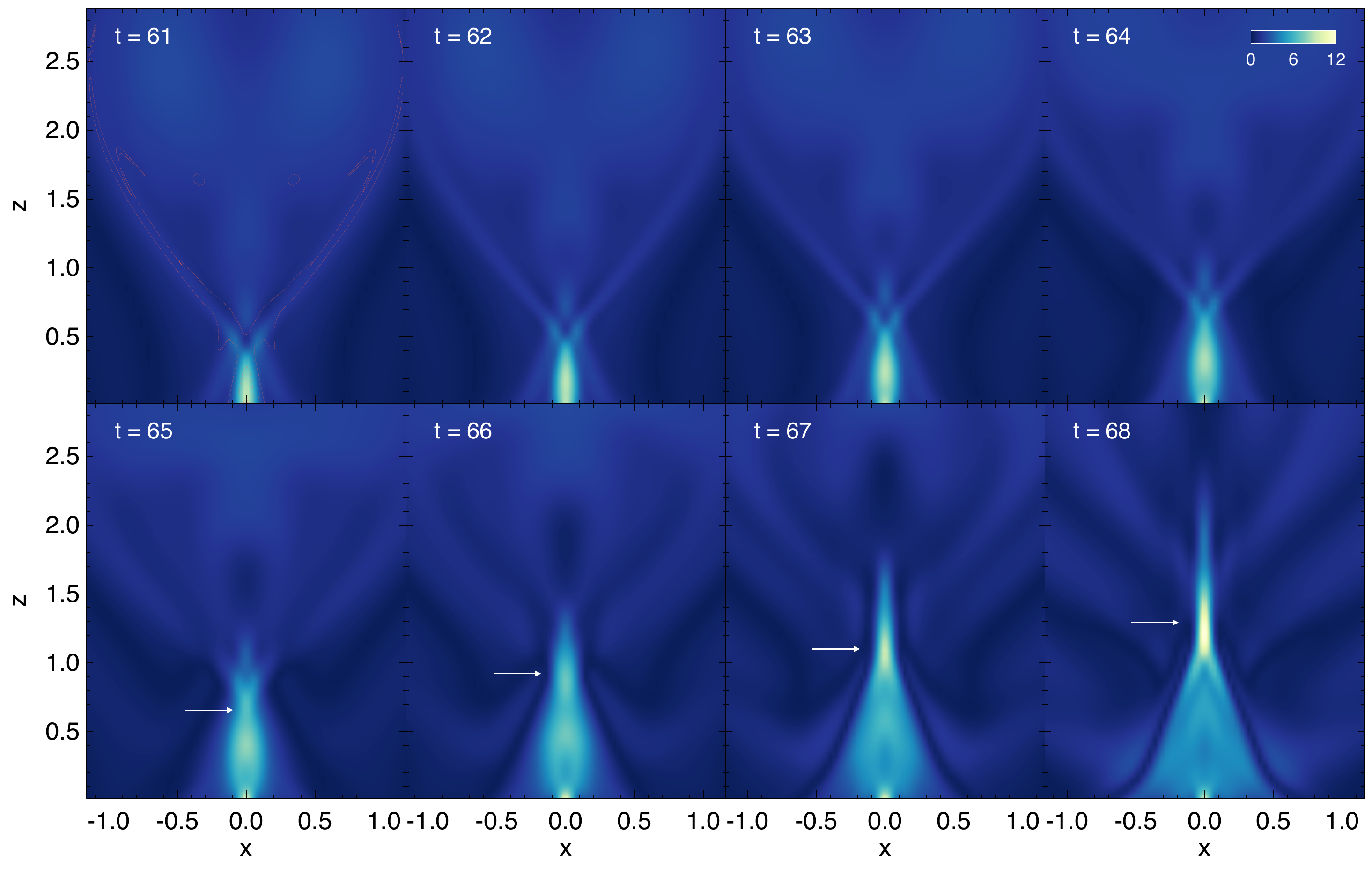}
\caption{\footnotesize{\textbf{Current structures in the reconnection region during $61\le t\le68$.} Each sub-panel shows the distribution of dimensionless $J/B$ on the plane $y=0$. The colorbar for $J/B$ in all sub-panels is displayed at the upper right corner. The red contours in first sub-panel show the outlines of QSLs at $t=61$. The white arrows mark the forming current sheet in the reconnection region.}}
\label{figs8}
\end{figure}

\clearpage
\newpage
\begin{figure}
\centering
\includegraphics[width=0.8\hsize]{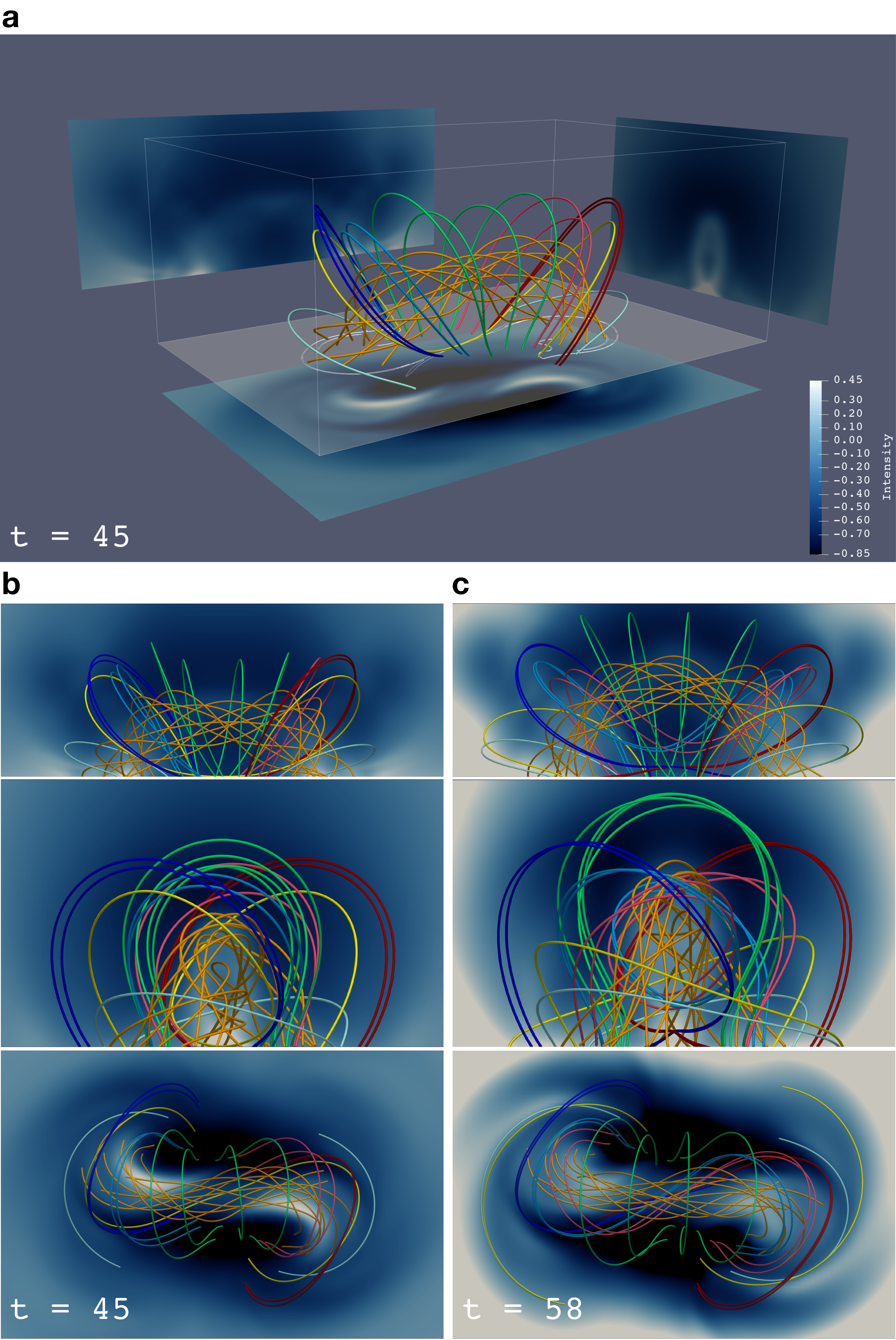}
\caption{\footnotesize{\textbf{Magnetic field lines and synthetic EUV images at the AIA 335 \AA.} (a) Same as Fig. \ref{fig1}a but for $t=45$. (b) Three side views of field lines and percentage difference images at $t=45$. (c) Same as the panel b but for $t=58$.}}
\label{figs9}
\end{figure}

\clearpage
\newpage
\begin{figure}
\centering
\includegraphics[width=\hsize]{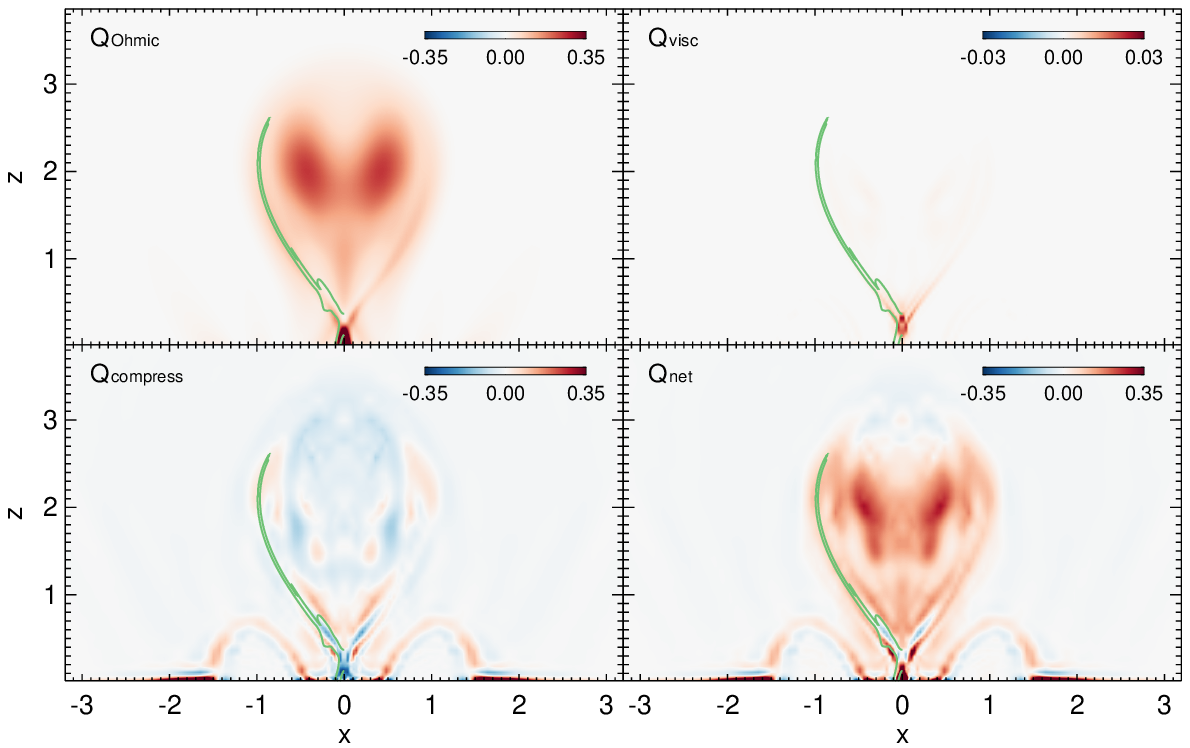}
\caption{\footnotesize{\textbf{Heating rates on the plane $y=0$ at $t=58$.} They are in order as follows: rates of Ohmic heating ($Q_{Ohmic}$), viscous heating ($Q_{visc}$), compression heating ($Q_{compress}$), and net heating ($Q_{net}$) which is the sum of the first three. All heating rates are in dimensionless unit. The field of view is the same as that of the synthetic image along the y-direction in Fig. \ref{fig1}a. The green contours represent those of QSLs at the half panel of $x\le0$.}}
\label{fig6}
\end{figure}

\clearpage
\newpage
\begin{figure}
\centering
\includegraphics[width=0.7\hsize]{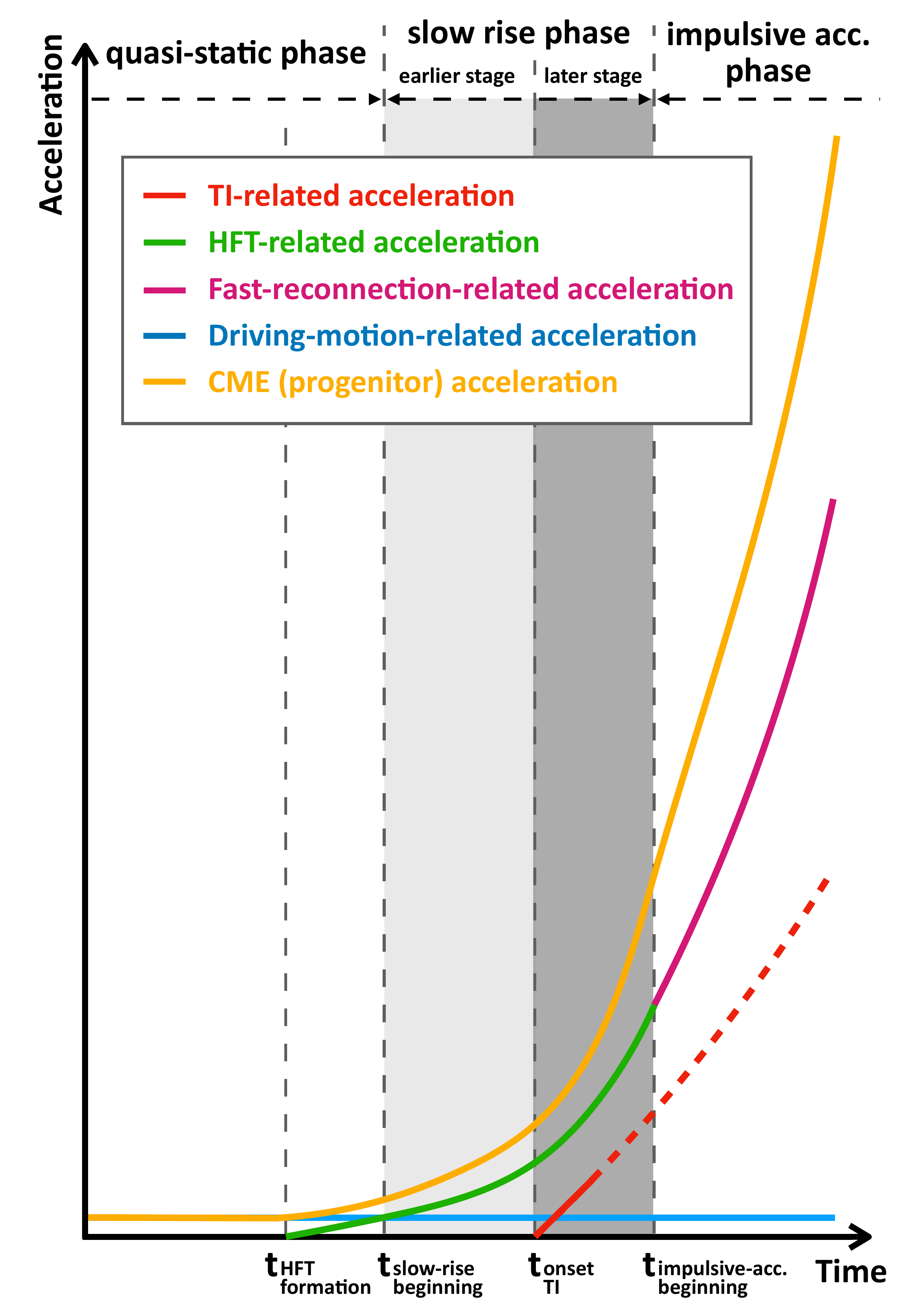}
\caption{\footnotesize{\textbf{Physical processes as disclosed by the kinematics of the CME progenitor and CME.} The yellow curve represents the acceleration of the CME (progenitor). The red, green, purple, blue curves show the four accelerations contributed by the torus instability, the HFT reconnection, the fast magnetic reconnection, and the driving motion, respectively. The onset times of the HFT, the slow rise, the torus instability, and the impulsive acceleration (abbreviated as acc.) of the CME are displayed by four dashed lines and labeled as $t_{\textup{HFT-formation}}$, $t_{\textup{slow-rise-beginning}}$, $t_{\textup{onset-TI}}$, and $t_{\textup{impulsive-acc.-beginning}}$, respectively. In our simulation, these four onset times correspond to $t=46$, $t=52$, a moment during the period $59\le t<63$, and $t=65$, respectively. The light grey and dark grey regions show the earlier stage and the later stage of the slow rise phase/CME initiation process, respectively. The red curve is dashed during a part of the slow rise phase and the impulsive acceleration  phase, indicating that the dominant mechanism for the acceleration is unknown whether it is torus instability or magnetic reconnection in these periods.}}
\label{fig5}
\end{figure}

\clearpage
\newpage
\appendix
\beginsupplement

\section{Analysis of Onset Time of Torus Instability}\label{AppA}
\subsection{Control Simulations}\label{AppA1}
In this work, we perform three control simulations to help us determine the onset time of the torus instability in ``Simulation De''. In the following, we first introduce the numerical setups in these three control simulations.

The first control simulation, named as ``Simulation Half-driven-eruption'' (abbreviated as ``Simulation He'') is composed of two phases: the shearing phase ($0\le t\le18$) and the converging phase ($18<t\le 70$). The setups in ``Simulation He'' are the same as those in ``Simulation De'', except that the amplitude of the converging flow is gradually switched off to half during $59<t\le60$ and then kept at half after that ($60<t\le70$) by modifying the function of $\gamma(t)$ in Equation \ref{eq10} in these periods:
\begin{equation}\label{eq13}
\begin{gathered}
\gamma(t)=\left\{
\begin{aligned}
-\frac{1}{4}\tanh[6.0(t-59.5)]+\frac{3}{4} & & & & & & 59<t\le60 \\
\frac{1}{2} & & & & & & 60<t\le70,
\end{aligned}
\right.
\end{gathered}
\end{equation}

The second control simulation, named as ``Simulation Undriven-eruption'' (abbreviated as ``Simulation Ue'') is composed of three phases: the shearing phase ($0\le t\le18$), the converging phase ($18<t\le 60$), and the relaxation phase ($60<t\le71$). Before and at $t=59$, the setups of ``Simulation Ue'' are the same as those of ``Simulation De''. After $t=59$, the converging flow is gradually switched off to zero during $59<t\le60$, and then the velocity on the cell-center bottom surface is fixed to zero during the relaxation phase ($60<t\le71$), also by modifying the function of $\gamma(t)$ in Equation \ref{eq10} in these periods:
\begin{equation}\label{eq14}
\begin{gathered}
\gamma(t)=\left\{
\begin{aligned}
-\frac{1}{2}\tanh[6.0(t-59.5)]+\frac{1}{2} & & & & & & 59<t\le60 \\
0 & & & & & & 60<t\le71.
\end{aligned}
\right.
\end{gathered}
\end{equation}
Correspondingly, during $59<t\le71$, the dissipation term in the Equation \ref{eq3} in the layer $k=1$ is modified into:
\begin{equation}\label{eq15}
\frac{\partial \boldsymbol{B}}{\partial t}+\nabla\cdot(\boldsymbol{v}\boldsymbol{B}-\boldsymbol{B}\boldsymbol{v}+\psi\boldsymbol{I}) = \left\{
\begin{aligned}
\eta(\frac{\partial^2}{\partial x^2}+\frac{\partial^2}{\partial y^2})\boldsymbol{B} & & & & & 59<t\le60 \\
0 & & & & & 60< t\le71.
\end{aligned}
\right.
\end{equation}
During $59<t\le60$, the setup of $\eta$ is the same as that during $18<t\le59$. During the relaxation phase ($60<t\le71$), $\eta$ is set to zero in the layer $k=1$ and set to be uniform ($\eta=4\times10^{-4}$; in dimensionless unit) in the whole region except the layer $k=1$. All other setups during $59<t\le71$ are the same as those before $t=59$.

The third control simulation, named as ``Simulation No-eruption'' (abbreviated as ``Simulation Ne'') is also composed of three phases: the shearing phase ($0\le t\le18$), the converging phase ($18<t\le 59$), and the relaxation phase ($59<t\le72$). The setups in three phases of ``Simulation Ne'' are the same as those in three phases of ``Simulation Ue'', respectively, except that the converging motion is switched off a little earlier, during $58<t\le59$.

\subsection{Kinematics of CME Progenitors/CMEs in Control Simulations}\label{AppA2}
The kinematics of CME progenitors and CMEs in these three control simulations is estimated by a method same as that applied to ``Simulation De''. For these three simulations, the point, from which the overlying field line is traced, is the same as that used in ``Simulation De''. The velocity at this fixed point is extremely small, very close to or equal to zero, in all phases of all three simulations.

The kinematics of CME progenitors and CMEs in control simulations and ``Simulation De'' is shown in Fig. \ref{figs4}a-c. We note that, for three control simulations, the decreasing acceleration or even negative acceleration in a period shortly after switching off the driving motion (e.g., $59\le t<61$ in ``Simulation He'', $59\le t<61$ in ``Simulation Ue'', $58\le t<62$ in ``Simulation Ne'') could be due to the viscous and resistive effects, as well as the relative relaxations of magnetic tension and pressure around the tipping point of the equilibrium curve \citep{Demoulin2010}.

\subsection{Onset Time of Torus Instability}\label{AppA3}
For all four simulations, in the period $57\le t\le64$, the flux rope is basically along the y-direction (Fig. \ref{figs3}; taking examples at $t=57$ and $t=64$ in ``Simulation De''). Therefore, the magnetic field at the intersection of the flux rope axis and the plane $y=0$ is considered in the y-direction in this period, and the intersection is determined with the contour of $B_x=0$ and the contour of $B_z=0$ in the plane $y=0$. The axis field line is then traced from the intersection. The decay index, $n=-\textup{d}(\textup{ln}B_t)/\textup{d}(\textup{ln}h)$, describes the decay rate of the component ($B_t$) of the potential field, which is transverse and perpendicular to the flux rope, with the height ($h$). At each analysed moment, the potential field is extrapolated by the Green's function method (performed by Solar Software (SSW; \citeauthor{Freeland2012}, \citeyear{Freeland2012}) package code ``optimization\_fff.pro'') with $B_z$ at the cell-center bottom surface as the input. During $57\le t\le64$, the component $B_t(h)$ is defined as the absolute value of $B_x$ of the potential field at $(0,0,h)$, considering the flux rope axis is along the y-direction and the horizontal coordinate of its apex is around $(0,0)$.

\begin{figure}
\centering
\includegraphics[width=\hsize]{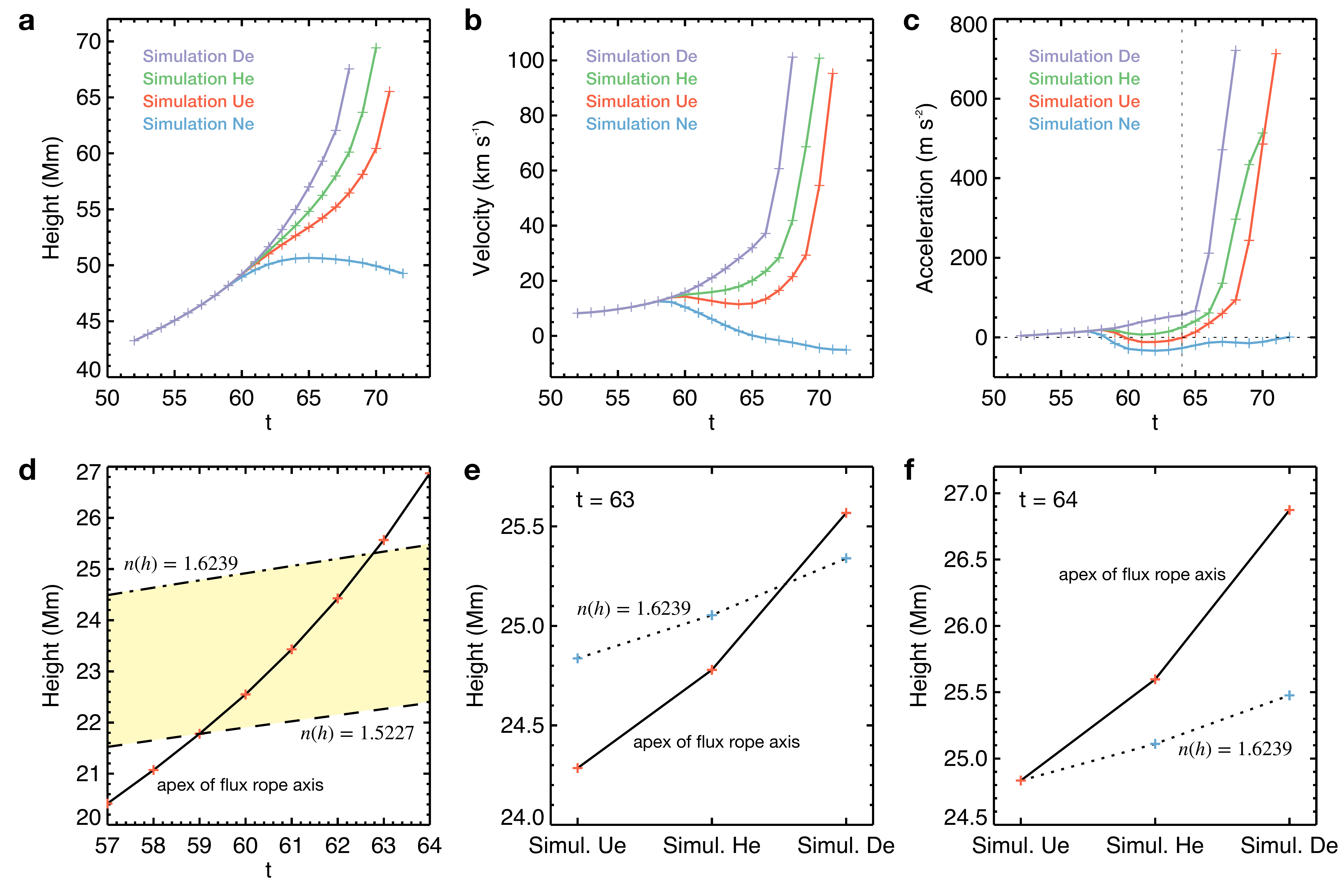}
\caption{\footnotesize{\textbf{Kinematics of CME progenitors and CMEs in four simulations and estimation of the onset time of the torus instability in ``Simulation De''.} (a)-(c) Kinematics of CME progenitors and CMEs in ``Simulation De'' (purple curves), ``Simulation He'' (green curves), ``Simulation Ue'' (red curves), and ``Simulation Ne'' (blue curves). In panel c, the vertical and horizontal black dashed lines mark that the flux rope acceleration in ``Simulation Ue'' is close to zero at $t=64$. (d) The solid curve exhibits the height of the flux rope axis apex during $57\le t\le64$ in ``Simulation De''. The dotted-dashed curve (long-dashed curve) represents the evolution of the height ($h$) during $57\le t\le64$ in ``Simulation De'', the height which meets the condition that the decay index at $(0,0,h)$ equals 1.6239 (1.5227) at each moment. The yellow region marks the range for the critical height of the torus instability onset during $57\le t\le64$ in ``Simulation De''. (e) The solid curve shows the evolution of the height of the flux rope axis apex at $t=63$, from ``Simulation (abbreviated as Simul.) Ue'' to ``Simulation He'' to ``Simulation De''. The dashed curve shows the evolution of the height ($h$), which meets the condition that the decay index at $(0,0,h)$ equals 1.6239 at $t=63$, from ``Simulation Ue'' to ``Simulation He'' to ``Simulation De''. (f) Same as panel e but for those at $t=64$.}}
\label{figs4}
\end{figure}

\begin{figure}
\centering
\includegraphics[width=\hsize]{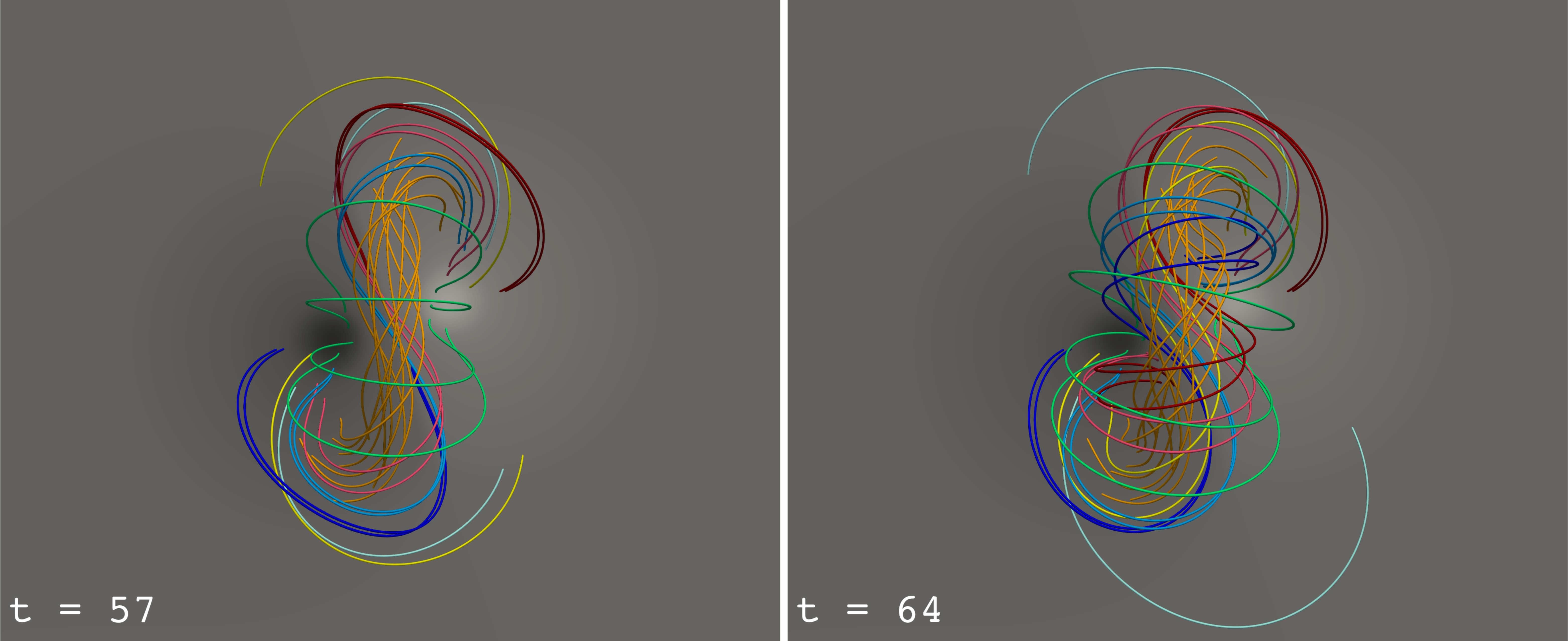}
\caption{\footnotesize{\textbf{Top views of magnetic field lines in Fig. \ref{fig1}b.} The left panel is  at $t=57$ and the right panel is at $t=64$. The vertical direction is the y-direction.}}
\label{figs3}
\end{figure}

We first analyze the kinematic evolution in ``Simulation Ue'' (see red curves in Fig. \ref{figs4}a-c). We consider that the onset time of the CME eruption in the physical sense (i.e., the time when the stable equilibria of the flux rope is broken) is between $t=59$ and $t=64$ for the following two reasons: (a) The CME progenitor fails to erupt in ``Simulation Ne'', in which all setups are the same as those in ``Simulation Ue'' but the converging motion is switched off during $58<t\le59$. This means that the CME in ``Simulation Ue'' erupts after $t=58$. The earliest limit of its eruption onset time is therefore set at $t=59$. (b) The acceleration in ``Simulation Ue'' changes from negative to positive at $t\sim64$ and increases continuously after that (Fig. \ref{figs4}c). This indicates that the flux rope is out of equilibrium after that, and thus the latest limit of the eruption onset time is set at $t=64$. For ``Simulation Ue'', the decay index of the potential field at the apex of the flux rope axis is 1.5227 at $t=59$ and 1.6239 at $t=64$, very close to the threshold of 1.5 for the occurrence of torus instability \citep{Kliem2006}, indicating that the CME eruption is highly probably triggered by the torus instability. The HFT reconnection that starts earlier ($t=46$) is considered unable to lead to the eruption as the eruption fails in the presence of HFT in ``Simulation Ne'' \citep{Aulanier2010}. 

Next, we analyze the kinematic evolution in ``Simulation De''. We consider that the (pre-)eruptive flux rope in ``Simulation De'' is quite similar to that in ``Simulation Ue'' in the period $59\le t\le64$, as the height of flux rope axis apex in ``Simulation Ue'' is only 7.6\% lower than that in ``Simulation De'' at $t=64$. Therefore, if the torus instability also sets in ``Simulation De'', the critical decay index for the torus instability onset should be in a range quite similar to $1.5227\le n\le1.6239$. In the following, we take the range $1.5227\le n\le1.6239$ for the critical decay index in ``Simulation De''. In Fig. \ref{figs4}d, it is clear that the height of the flux rope axis apex in ``Simulation De'' is already higher than the height of the upper limit of the critical decay index at $t=63$, indicating that the torus instability could set in ``Simulation De'' and that its onset time should be before $t=63$. The earliest onset time of the torus instability in ``Simulation De'' is still $t=59$, as the apex of the flux rope axis reaches the height of the lower limit of critical decay index at this moment. Therefore, we summarize that the torus instability sets in ``Simulation De'' during $59\le t<63$.

We note that it is not an accident that the time range of the torus instability onset in ``Simulation De'' is earlier than that in ``Simulation Ue''. In the period $60< t\le64$, the driving motion becomes stronger and stronger from ``Simulation Ue'' to ``Simulation He'' to ``Simulation De'', with the maximum speed of the driving motion varying from 0 to 0.08 to 0.16 (in dimensionless unit). In Fig. \ref{figs4}e,f, we show that, on the one hand, for these three simulations, the flux rope in the simulation with a stronger driving motion could rise to a higher altitude at the same moment (taking examples at $t=63$ (see panel e) and at $t=64$ (see panel f)). On the other hand, in comparison among the three simulations, the one with a stronger driving motion has a higher height corresponding to the same critical decay index at the same moment (see panels e and f). However, it is clear that such a difference in the height of the flux rope axis is much larger than the difference in the height of a certain decay index, when comparing each two of these three simulations. This suggests that in comparison among different simulations, the stronger driving motion imposed, the earlier the flux rope axis reaches the height of the critical decay index and thus the earlier the torus instability starts, if regardless of the change in the critical decay index itself.

\section{Flux Rope Boundary}\label{AppB}
The flux rope boundary and the reconnection region are identified with the squashing degree $Q$ which measures the mapping of the field lines. The squashing degree $Q$ is derived by \citep{Titov2002}:
\begin{equation}\label{eq16}
Q=\frac{(\frac{\partial X}{\partial x})^2+(\frac{\partial X}{\partial y})^2+(\frac{\partial Y}{\partial x})^2+(\frac{\partial Y}{\partial y})^2}{\vert\frac{\partial X}{\partial x}\frac{\partial Y}{\partial y}-\frac{\partial X}{\partial y}\frac{\partial Y}{\partial x}\vert}
\end{equation}
with the code FastQSL \citep{Zhang2022}, where $(x,y)$ and $(X,Y)$ are coordinates of two footpoints of a field line.

\begin{figure}
\centering
\includegraphics[width=\hsize]{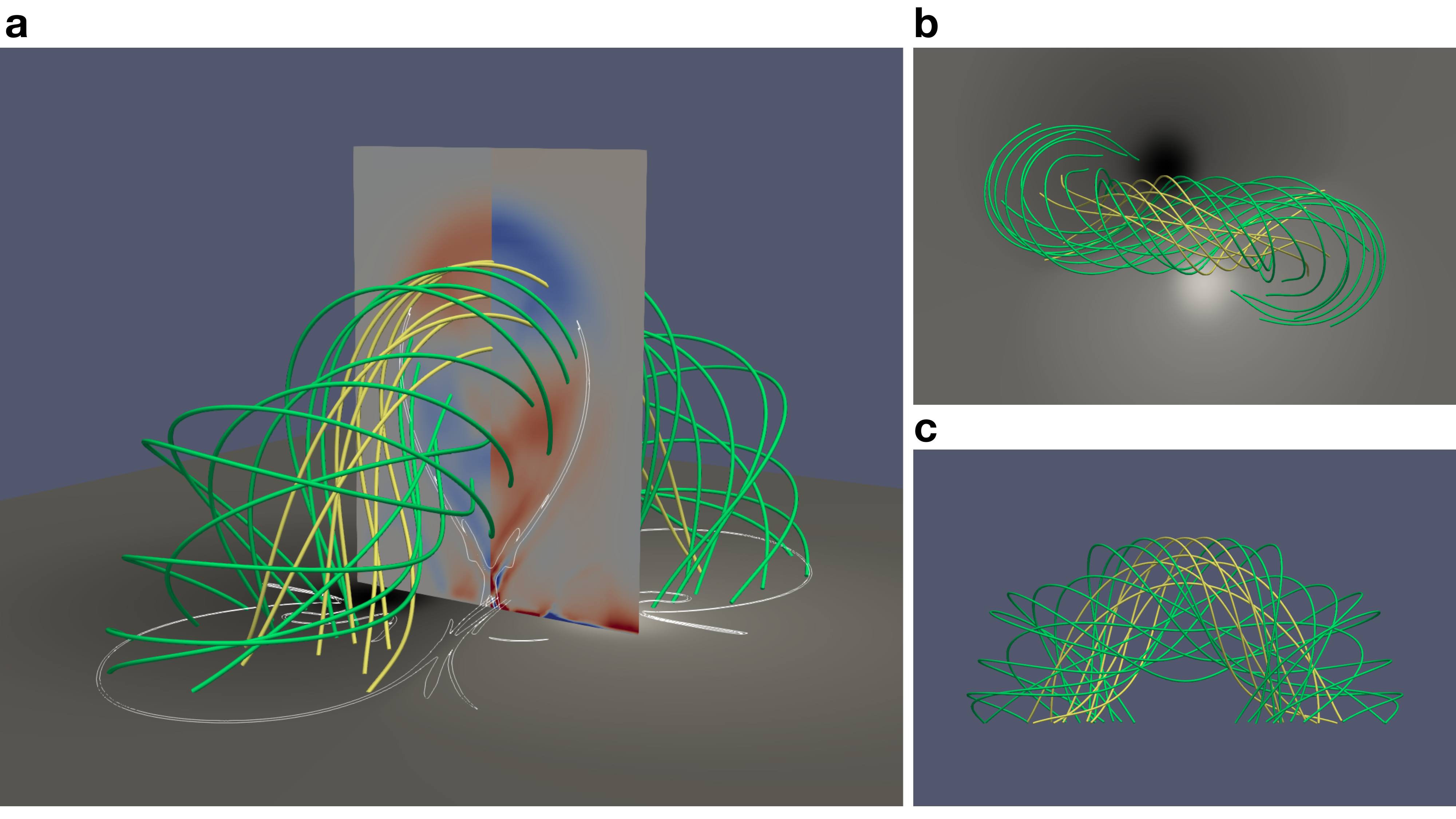}
\caption{\footnotesize{\textbf{Magnetic field lines in the flux rope at $t=58$.} (a) Forces on the plane $y=0$ and field lines in the flux rope. The green (yellow) shows the highly (weakly) twisted field lines. In the plane $y=0$, the left half panel shows the z-component of the thermal pressure gradient force and the right half panel shows the z-component of the Lorentz force. The colors and the scales of the forces are the same as those in Fig. \ref{figs5}. The cell-center bottom surface shows the distribution of $B_z$. The white contours on the bottom surface and the vertical plane represent the contours of QSLs. (b) Top view of panel a. (c) Face-on view of panel a.}}
\label{figs6}
\end{figure}

The QSLs refer to the locations where $\textup{log}Q$ is much larger than 2 \citep{Titov2002}, and here we define the QSLs as the region where $\textup{log}Q\ge3$. In the following we take the flux rope at $t=58$ in ``Simulation De'' as an example to explain how we determine the flux rope boundary. In the plane $y=0$, the bottom and side boundaries of the flux rope can be exactly determined by the QSLs, while the top boundary is blurred as the QSLs are not closed there (Fig. \ref{figs6}). We determine the top boundary qualitatively at the top edge of the strong downward Lorentz force region, i.e., the apex of the flux rope is roughly at $z=3.3$ (in dimensionless unit; see Fig. \ref{figs5}a). This determination is reasonable for the following reasons. First, as shown in Fig. \ref{figs6}, the weakly twisted field lines traced from the downward Lorentz force region (yellow field lines) are mostly anchored in the regions partially surrounded by the QSL footprints. They compose a flux rope configuration together with the highly twisted field lines traced from the upward Lorentz force region (green field lines). Second, the tops of the high mass density region (represented by the gravity in Fig. \ref{figs5}c) and of the high current density region (represented by the Ohmic heating rate in Fig. \ref{fig6}) are also around $z=3.3$ in the plane $y=0$, also indicating that the flux rope apex is around $z=3.3$.

\section{Estimating the Reconnection Electric Field}\label{AppC}
We estimate the reconnection electric field with a method commonly used in observations \citep{Qiu2002}. In brief, the reconnection electric field ($E$) is derived by $E=v_QB_z$, where $v_Q$ is the separation speed of the QSL footprint in the direction perpendicular to the PIL and $B_z$ is the z-component of the magnetic field at the QSL footprint. This estimation method is based on a 2D flare model but is considered still applicable to a 3D situation \citep{Forbes2000}. We refer readers to the previous paper \citep{Qiu2002} for more details of this method.

In Fig. \ref{fig4}a-e, we show the separation motion of the QSL footprint on the cell-center bottom surface. To estimate the separation motion speed, we set a slit (represented by the orange dashed lines in Fig. \ref{fig4}a-c) on this surface, starting from the origin and along the direction perpendicular to the PIL. The time-slice plot of the normalized $\textup{log}Q$ (Fig. \ref{fig4}f), where the $\textup{log}Q$ on the slit is normalized by its maximum at each moment, exhibits how the QSL footprint separates from the PIL. The center of the QSL footprint, where the normalized $\textup{log}Q$ reaches its local maximum, is marked by the red symbol at each moment. The phenomenon that the QSL footprint stays on the PIL before and at $t=45$ confirms that the magnetic topology of the reconnection region is a BP in this period. The separation of QSL footprint starts at $t=46$, marking the first appearance of HFT at this moment. The separation speed ($v_Q$) is derived to be the derivative of the distance from the origin to the QSL footprint with time. Finally, the electric field of the HFT reconnection is derived by multiplying the separation speed with the local $B_z$. The electric field during $46\le t\le51$ (see Fig. \ref{fig4}g,h) is represented by that derived from the orange slit, considering the reconnection still occurs in BP rather than HFT in many other places in this period. The electric field during $52\le t\le68$ (also see Fig. \ref{fig4}g,h) is represented by the average of the electric fields derived from the orange slit and three other slits (represented by the pink dashed lines in Fig. \ref{fig4}a-c; all of them crossing through HFT footprints rather than BP footprints since $t=52$). The aim of the multiple measurements and averaging is to better show the evolution of the electric field in the entire HFT.

\section{Synthetic EUV Images of the CME Progenitor}\label{AppD}
We synthesize the EUV images at the AIA 335 \textup{\AA} as observed from three side views through integrating the emissivity along each direction under the optically thin emission assumption. The passband is selected at 335 \textup{\AA} to best demonstrate the structure of the hot CME progenitor. The emissivity is derived from the temperature and the number density of electron (which are dimensionalized with the dimensionless units; the number density is derived under the assumption of a fully ionized ideal gas with a hydrogen-helium abundance ratio of 10:1) with the AIA response function (performed by SSW package code ``aia\_get\_response.pro''), and the integration is performed in a box where $-3.2\le x\le3.2$, $-4.9\le y\le4.9$, and $0.015\le z\le3.86$ (in dimensionless unit). To better show the flux rope structure, we show the percentage difference intensity rather than the original synthetic intensity, which is derived by:
\begin{equation}\label{eq17}
D_i=\frac{I_i-I_0}{I_0},
\end{equation}
where $D_i$ and $I_i$ are the percentage difference intensity and the original synthetic intensity at $t=t_i$, respectively, and $I_0$ is the original synthetic intensity at $t=0$. We note that the percentage difference intensity in the flux rope, although sometimes negative, is still larger than that of its surrounding, clearly showing the presence of the hot CME progenitor.

\clearpage
\newpage

\clearpage
\newpage

\clearpage
\newpage

\end{document}